\documentclass[pra,twocolumn,aps,showpacs,superscriptaddress]{revtex4}

\usepackage{amsmath}
\usepackage{amssymb}
\usepackage{hyperref}
\usepackage{graphicx}
\usepackage{subfigure}
\usepackage[usenames,dvipsnames]{color}
\usepackage{setspace}

\begin{document}

%%%%%%%%%%%%%%%%%%%%%%%%%%%%%%%%%%%%%%%%%%%%%%%%%%%%%%%%%%%%%%%%%%%%%%%%%%%%%%
%%%%                     Title and authors                                %%%%
%%%%%%%%%%%%%%%%%%%%%%%%%%%%%%%%%%%%%%%%%%%%%%%%%%%%%%%%%%%%%%%%%%%%%%%%%%%%%%

\title{Dynamics of the corotating vortices in dipolar Bose-Einstein 
       condensates in the presence of dissipation}

\author{S. Gautam}
\affiliation{Instituto de F\'{\i}sica Te\'orica, UNESP - Universidade Estadual
             Paulista, \\ 01.140-070 S\~ao Paulo, S\~ao Paulo, Brazil}

%%%%%%%%%%%%%%%%%%%%%%%%%%%%%%%%%%%%%%%%%%%%%%%%%%%%%%%%%%%%%%%%%%%%%%%%%%%%%%
%%%%%%%%%%                    Abstract                             %%%%%%%%%%%
%%%%%%%%%%%%%%%%%%%%%%%%%%%%%%%%%%%%%%%%%%%%%%%%%%%%%%%%%%%%%%%%%%%%%%%%%%%%%%

\date{\today}
\begin{abstract}
 We study the dynamics of a single and a corotating vortex pair in a dipolar
Bose-Einstein condensate in the framework of dissipative Gross-Pitaevskii 
equation. This simple model enables us to simulate the effect of finite 
temperature on the vortex dynamics. We study the effect of dipolar 
interactions on the dynamics of a single vortex in the presence of 
phenomenological dissipation. In the case of a corotating vortex pair, an 
initial asymmetry in the locations of the vortices can lead to different 
decay rates for the constituent vortices as is the case for the condensates
interacting via pure contact interactions. We observe that the anisotropic 
interaction between the component vortices manifests itself as the 
perceptible difference in the trajectories traversed by the vortices in the 
condensate at finite temperatures. 
\end{abstract}


\maketitle

%%%%%%%%%%%%%%%%%%%%%%%%%%%%%%%%%%%%%%%%%%%%%%%%%%%%%%%%%%%%%%%%%%%%%%%%%%%%%%
%%%%%%%%%%%%%                 Introduction                         %%%%%%%%%%%
%%%%%%%%%%%%%%%%%%%%%%%%%%%%%%%%%%%%%%%%%%%%%%%%%%%%%%%%%%%%%%%%%%%%%%%%%%%%%%

\section{Introduction}
\label{I}
 Quantized vortices in the Bose-Einstein condensates (BECs), which unequivocally 
prove the superfluid nature of these systems, can be produced experimentally
using a variety of methods \cite{Matthews,Madison,Raman,Abo-Shaeer,Leanhardt}.
Besides a single vortex \cite{Matthews,Madison} and a vortex lattice 
consisting of corotating vortices \cite{Abo-Shaeer}, creation of the 
vortex-antivortex pairs or vortex dipoles has also been observed during (a) 
the motion of the condensate across the Gaussian obstacle potential 
\cite{Neely}, (b) the decay of the dark soliton in the quasi two-dimensional 
condensates \cite{Anderson}, and (c) the rapid quench of the condensate through
the condensation temperature \cite{Weiler,Freilich}. Freilich {\em et al.}
have performed the experimental study of the real time dynamics of a single
vortex as well as a vortex dipole \cite{Freilich}. In a recent experiment,
the dynamics of $2$-$4$ corotating vortices was also studied \cite{Navarro}.
At zero temperature, the dynamics of vortex dipoles has been investigated 
theoretically in Refs. \cite{Middelkamp,Torres,Kuopanportti}. The generation 
and energetic stability of the vortex dipoles in phase-separated binary 
condensates was also theoretically investigated in a couple of our earlier 
works \cite{Gautam-1}. At finite temperature, an off-centered vortex in a 
non-rotating condensate moves out of the condensate following a spiral 
trajectory, and it has been confirmed by various theoretical investigations 
\cite{Schmidt,Madarassy,Jackson,Allen,Wright,Rooney,Duine-2,Yan}. 
Recently we studied the dynamics of a single vortex as well a vortex pair in
the BECs at finite temperature using the stochastic Gross-Pitaevskii equation 
(SGPE) \cite{Gautam}. 

 All the studies mentioned in the previous paragraph have dealt with the 
BECs interacting via short range and isotropic contact interactions. Dipolar 
interactions were not considered in any of these studies. The primary 
motivation of the present work is to study the dynamics of a single and a 
pair of corotating vortices in the dipolar Bose-Einstein condensates (DBECs) 
at finite temperatures. A lot of experimental and theoretical studies have 
been done on the dipolar quantum gases since the first experimental 
realization of the Bose-Einstein condensation in a gas of chromium 
($^{52}$Cr) atoms, which have permanent magnetic dipole moments 
\cite{Griesmaier,Lahaye-1,Koch}. In addition to $^{52}$Cr, Bose-Einstein 
condensation in the systems with stronger dipolar interactions like 
dysprosium ($^{164}$Dy) \cite{Lu} and erbium ($^{168}$Eb) \cite{Aikawa} has 
also been observed experimentally. The various developments in the field of 
the dipolar quantum degenerate gases have been reviewed in 
Refs. \cite{Lahaye-2, Baranov}. Dipolar interaction is characterized by its 
long range and anisotropic nature, which is not the case with contact 
interactions. Also, the inter-particle interaction in dipolar condensates is 
momentum dependent as all the partial waves contribute to the scattering 
amplitude \cite{Lahaye-2}. This leads to the emergence of roton like 
excitations in the system \cite{Santos,Wilson-1, Wilson-2, Fischer}. The presence of 
roton like excitations is responsible for variety of effects like lowering of
the Landau critical speed \cite{Landau} below the speed of sound for 
sufficiently large particle number \cite{Wilson-2}, density fluctuations at 
defects like vortices \cite{Wilson-1,Yi}, roton instability 
\cite{Santos,Ronen,Wilson-3,Fischer}, enhancement of the density fluctuations in 
two-dimensional DBECs \cite{Boudjemaa}, and anisotropic superfluidity 
\cite{Ticknor,Muruganandam} which in turn leads to the anisotropic 
merging and splitting in quasi two-dimensional DBECs \cite{Gautam-3}. At 
zero temperature, the effect of dipolar interactions and the relative 
strength of dipolar and contact interactions on the structure and stability 
of the vortices in the DBECs has been theoretically investigated 
\cite{Wilson-1,Yi,Abad}. The dynamical instability of the rotating dipolar 
condensates, which can lead to vortex lattice formation, has been studied in 
Ref.~\cite{Bijnen}. The vortex lattice formation in the rotating quasi two
dimensional DBECs has also been studied theoretically \cite{Malet,Kumar}.
  
In the present work, we study the dynamics of a single and a corotating 
vortex pair in the framework of dissipative Gross-Pitaevskii equation (DGPE).
The DGPE is the simplest model that can simulate the effect of the finite 
temperature on the vortex dynamics in  
DBECs \cite{Pitaevskii}. The microscopic origin of the dissipation in
the non-dipolar condensates at finite temperature condensates has been 
discussed in Refs. \cite{Penckwitt, Blakie}.

The paper is organized as follows- We provide the mean-field description of 
DBECs in Sec.~\ref{Sec-II} in the presence of phenomenological dissipation. 
Here we discuss the quasi two-dimensional non-local dissipative 
Gross-Pitaevskii equation (DGPE) which we employ to study the 
dipolar Bose-Einstein condensate (DBEC) with an 
arbitrary direction of polarization. In Sec.~\ref{Sec-III}, we study the 
dynamics of a single and a corotating vortex pair in the presence of 
dissipation. We conclude by providing a summary of results in 
Sec.~\ref{Sec-IV}.

%%%%%%%%%%%%%%%%%%%%%%%%%%%%%%%%%%%%%%%%%%%%%%%%%%%%%%%%%%%%%%%%%%%%%%%%%%%%%%
%%%%%%%%%           Mean Field Description                         %%%%%%%%%%%
%%%%%%%%%%%%%%%%%%%%%%%%%%%%%%%%%%%%%%%%%%%%%%%%%%%%%%%%%%%%%%%%%%%%%%%%%%%%%%
\section{Mean Field Description of Dipolar Bose-Einstein condensate}
\label{Sec-II}
In the mean field regime, a dipolar Bose-Einstein condensate (DBEC) at 
finite temperature can be modeled by the non-local DGPE
\cite{Lahaye-2, Baranov,Madarassy}
\begin{eqnarray}
i\hbar\frac{\partial \Psi(\mathbf x,t)}{\partial t} &=&
(1-i\gamma)\left[-\frac{\hbar^2\nabla^2}{2m}
+ V(\mathbf x) +g|\Psi(\mathbf x,t)|^2 +\right.\nonumber\\
 &&\left.\int U_{\rm dd}(\mathbf{x-x'})
|\Psi(\mathbf x',t)|^2 d\mathbf x'\right] \Psi(\mathbf x,t),
\label{Eq.1}
\end{eqnarray}
where $\Psi(\mathbf x, t)$ is the wave function of the condensate and 
$\gamma$ is the dissipation introduced by the thermal cloud. The last term 
in the square brackets accounts for the long range dipole-dipole interaction.
When all the dipoles point in the same direction, i.e., the dipolar gas is 
polarized, the dipole-dipole interaction strength is
\begin{equation}
U_{\rm dd} = \frac{C_{\rm dd}}{4\pi}
             \frac{1-3\cos^2\theta}{|\mathbf x-\mathbf x'|^3},
\end{equation}
where $\theta$ is the angle between the direction of polarization and 
relative position vector of the dipoles. The coupling constant 
$C_{\rm dd} = 12\pi\hbar^2a_{\rm dd}/m$, where $a_{\rm dd}$ is the length
characterizing the strength of the dipolar interactions and $m$ is the mass 
of the atom. For the dipolar gas consisting of atoms with permanent magnetic 
dipole moment $\chi$ like 
$^{52}$Cr, $a_{\rm dd} = \mu_0\chi^2m/(12\pi\hbar^2)$, where $\mu_0$ is the
permeability of the free space. In the present work, we consider the DBEC 
trapped in a harmonic trapping potential 
$V (\mathbf x) = m(\omega_x^2 x^2 + \omega_y^2 y^2 + \omega_z^2 z^2 )/2$,
where $\omega_j$'s with $j=x,y,z$ are trapping
frequencies along the three coordinate axes. The interaction strength 
$g = 4\pi\hbar^2 a/m$, where $a$ is the $s$-wave scattering length,
characterizes the contact interaction between atoms. The total number of 
atoms $N$ and energy
$E$ are not conserved by Eq.~(\ref{Eq.1}) due to the presence of dissipation. 
For the sake of solving Eq.~(\ref{Eq.1}) numerically, we transform the DGPE 
into dimensionless form using following transformations:
\begin{eqnarray}
\mathbf x &= &\tilde{\mathbf {x}} a_{\rm osc},~t  =
\frac{2\tilde {t} \omega^{-1}}{1+\gamma^2},~
\Psi = \sqrt{N}\frac{\tilde {\Psi}}{a_{\rm osc}^{3/2}},
\end{eqnarray}
where $a_{\rm osc} = \sqrt{\hbar/(m\omega)}$ with $\omega =
\rm min~ \{\omega_x,\omega_y,\omega_z\}$ is the oscillator length.
The dimensionless DGPE for the DBEC is now of the 
form
\begin{eqnarray}
(i-\gamma)\frac{\partial \tilde{\Psi}}{\partial \tilde{t}}
&=& \left[-\tilde{\nabla}^2 + 2\tilde{V} +2\tilde{g}|\tilde{\Psi}|^2
+\right.\nonumber\\
&& \left. 2\int \tilde{U}_{\rm dd}(\tilde{\mathbf{x}} - \tilde{\mathbf{x'}})|\tilde{\Psi}
(\tilde{\mathbf{x'}})|^2d\tilde{\mathbf{x'}}\right] \tilde{\Psi},
\label{Eq.4}
\end{eqnarray}
where $\tilde V = (\lambda_x^2\tilde{x}^2+\lambda_y^2\tilde{y}^2
+\lambda_z^2\tilde{z}^2)/2$, $\lambda_j = \omega_j/\omega$
with $j = x,y,z$ and $\tilde g = 4\pi\hbar a N/(m\omega a_{\rm osc}^3)$.
 In the present work, we consider the DBEC in a quasi two-dimensional
trap with $\lambda_x=\lambda_y=1\ll\lambda_z$. In this case, the axial
degrees of freedom of the system are frozen. We write
$\tilde{\Psi}(\tilde{\mathbf x}) = \tilde{\zeta}(z)\tilde{\psi}(\tilde{x},\tilde{y})$ with
$\tilde{\zeta}(\tilde{z}) = (\lambda_z/\pi)^{1/4}e^{-(\lambda_z \tilde{z}^2)/2}$ as the harmonic
oscillator ground state along the axial direction. After integrating out the
axial coordinate, we obtain the following two-dimensional equation
\cite{Pedri, Muruganandam-1,Fischer,Ticknor} in dimensionless form:
\begin{eqnarray}
(i-\gamma)\frac{\partial \tilde{\psi}}{\partial \tilde{t}} &=& \left\{-\tilde{\nabla}_{\tilde{r}}^2
+ 2\tilde{V}_{\tilde{r}} +2\tilde{g}_{\tilde{r}}|\tilde{\psi}|^2 +4\sqrt{2\pi\lambda_z}
\tilde{a}_{\rm dd}N\right.
\nonumber\\
&& \times\int \frac{d^2 \tilde{k}_{\tilde{r}}}{4\pi^2}
e^{i \tilde{\mathbf{k}}_{\tilde{r}}.\tilde{\mathbf{r}}} \tilde{n}(\tilde{\mathbf k}_{\tilde{r}})
\left[\cos^2(\alpha)h_{2d}^{\perp}\left(\frac{\tilde{\mathbf{k}}_{\tilde{r}}}
{\sqrt{2\lambda_z}}\right) \right.\nonumber\\
&&\left.\left.+\sin^2(\alpha)h_{2d}^{\parallel}\left(\frac{\mathbf{\tilde{k}}_{\tilde{r}}}
{\sqrt{2\lambda_z}}\right)\right]\right\} \tilde{\psi},
\label{gpe_scaled}
\end{eqnarray}
where $\tilde{\nabla}_{\tilde{r}}^2 = \partial^2/\partial \tilde{x}^2+ \partial^2/\partial \tilde{y}^2$,
$\tilde{V}_{\tilde{r}} = \tilde{x}^2/2+\tilde{y}^2/2$, 
$\tilde{g}_{\tilde{r}} = \sqrt{\lambda_z/2\pi}\tilde{g}$, and $\tilde{a}_{\rm dd} = a_{\rm dd}/a_{\rm osc}$. Here we
have considered an arbitrary direction of polarization in the $xz$ plane, 
which makes an angle $\alpha$ with the $z$ axis, and
$h_{2d}^{\perp}(\tilde{\mathbf{k}}_{\tilde{r}}/\sqrt{2\lambda_z})$ and
$h_{2d}^{\parallel}(\tilde{\mathbf{k}}_{\tilde{r}}/\sqrt{2\lambda_z})$ are defined as
\begin{eqnarray}
h_{2d}^{\perp}\left(\frac{\tilde{\mathbf{k}}_{\tilde{r}}}{\sqrt{2\lambda_z}}\right) & =
&2-\frac{3\sqrt{\pi}\tilde{k}_{\tilde{r}}}{\sqrt{2\lambda_z}}
\exp\left(\frac{\tilde{k}_{\tilde{r}}^2}{2\lambda_z}\right){\rm erfc}\left(\frac{\tilde{k}_{\tilde{r}}}
{\sqrt{2\lambda_z}}\right),\nonumber\\
h_{2d}^{\parallel}\left(\frac{\tilde{\mathbf{k}}_{\tilde{r}}}{\sqrt{2\lambda_z}}\right) & =
& -1+\frac{3\sqrt{\pi}\tilde{k}_{\tilde{x}}^2}{\sqrt{2\lambda_z}\tilde{k}_{\tilde{r}}} 
\exp\left(\frac{\tilde{k}_{\tilde{r}}^2}{2\lambda_z}
\right){\rm erfc}\left(\frac{\tilde{k}_{\tilde{r}}}{\sqrt{2\lambda_z}}\right),\nonumber
\end{eqnarray}
where $\tilde{k}_{\tilde{r}} = k_ra_{\rm osc}$ and $\tilde{k}_{\tilde{x}} = k_xa_{\rm osc}$.
The scaled wavefunction is normalized to unity, i.e., 
$\int|\tilde{\psi}|^2d\tilde{x}d\tilde{y} = 1$.
For the sake of notational simplicity, from here on we will write the
scaled variables without tildes unless stated
otherwise.
We use the time splitting Fourier spectral method to solve equation
Eq.~(\ref{gpe_scaled}) \cite{Bao}. The spatial and time step sizes employed in
the present work are $0.11~a_{\rm osc}$ and $0.0012~{\omega^{-1}}$
respectively. The wavefunction is normalized to unity after each iteration in
time as it is no longer conserved in the presence of dissipation.

%%%%%%%%%%%%%%%%%%%%%%%%%%%%%%%%%%%%%%%%%%%%%%%%%%%%%%%%%%%%%%%%%%%%%%%%%%%%%%%
%%%%%%%%%%%%               Numerical study                         %%%%%%%%%%%%
%%%%%%%%%%%%%%%%%%%%%%%%%%%%%%%%%%%%%%%%%%%%%%%%%%%%%%%%%%%%%%%%%%%%%%%%%%%%%%%

\section{Dynamics of vortices}
\label{Sec-III}
\begin{center}
\begin{figure}[!h]
\begin{tabular}{c}
\resizebox{!}{!}
{\includegraphics[trim = 1.5cm 0cm 1.5cm 2cm,clip, angle=0,width=4cm]
                 {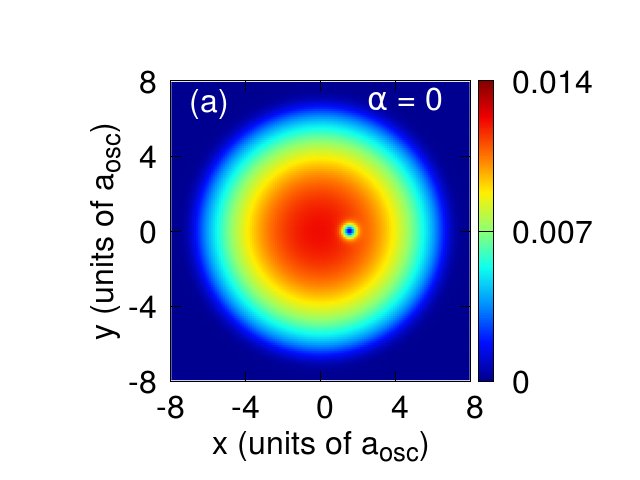}}
\resizebox{!}{!}
{\includegraphics[trim = 1.5cm 0cm 1.5cm 2cm,clip, angle=0,width=4cm]
                 {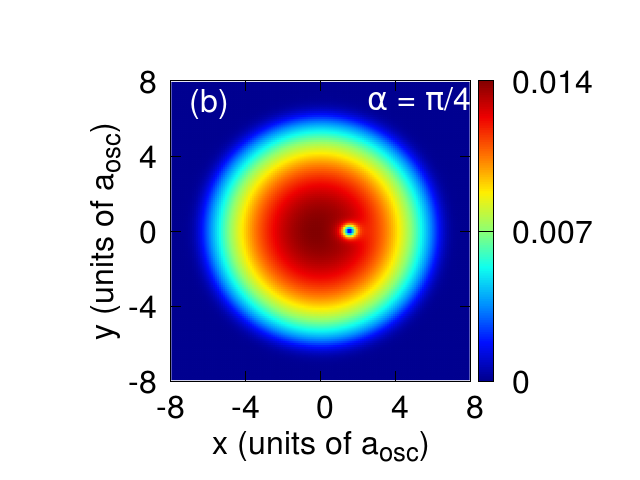}}\\
\resizebox{!}{!}
{\includegraphics[trim = 1.5cm 0cm 1.5cm 2cm,clip, angle=0,width=4.0cm]
                 {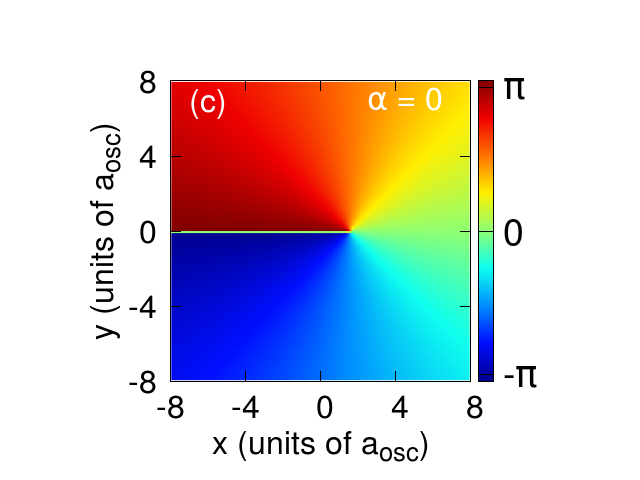}}
\resizebox{!}{!}
{\includegraphics[trim = 1.5cm 0cm 1.5cm 2cm,clip, angle=0,width=4.0cm]
                 {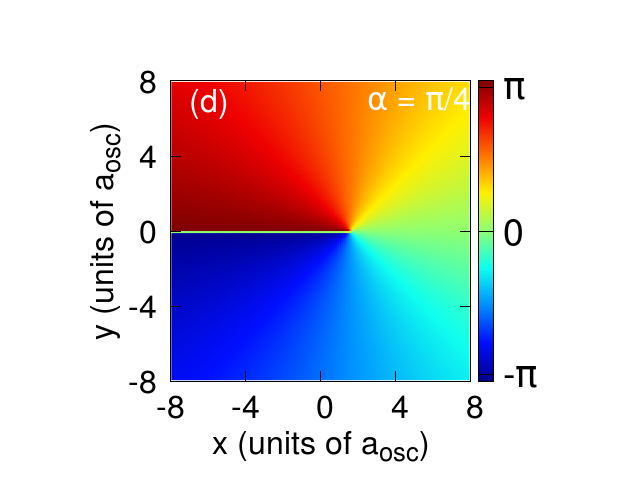}}\\
\resizebox{!}{!}
{\includegraphics[trim = 1.2cm 0cm 1.2cm 0cm,clip, angle=0,width=4.0cm]
                 {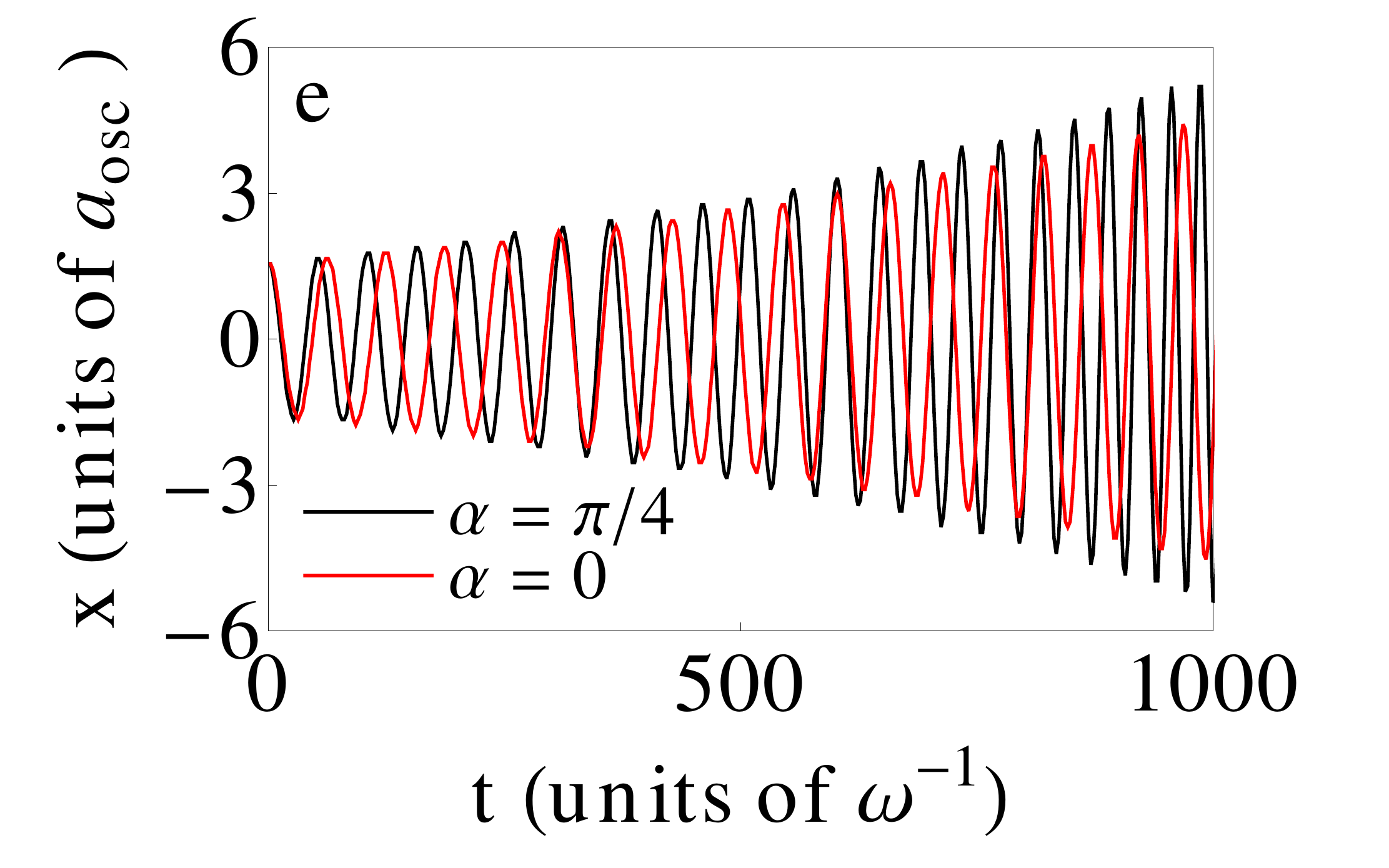}}
\resizebox{!}{!}
{\includegraphics[trim = 1.2cm 0cm 1.2cm 0cm,clip, angle=0,width=4.0cm]
                 {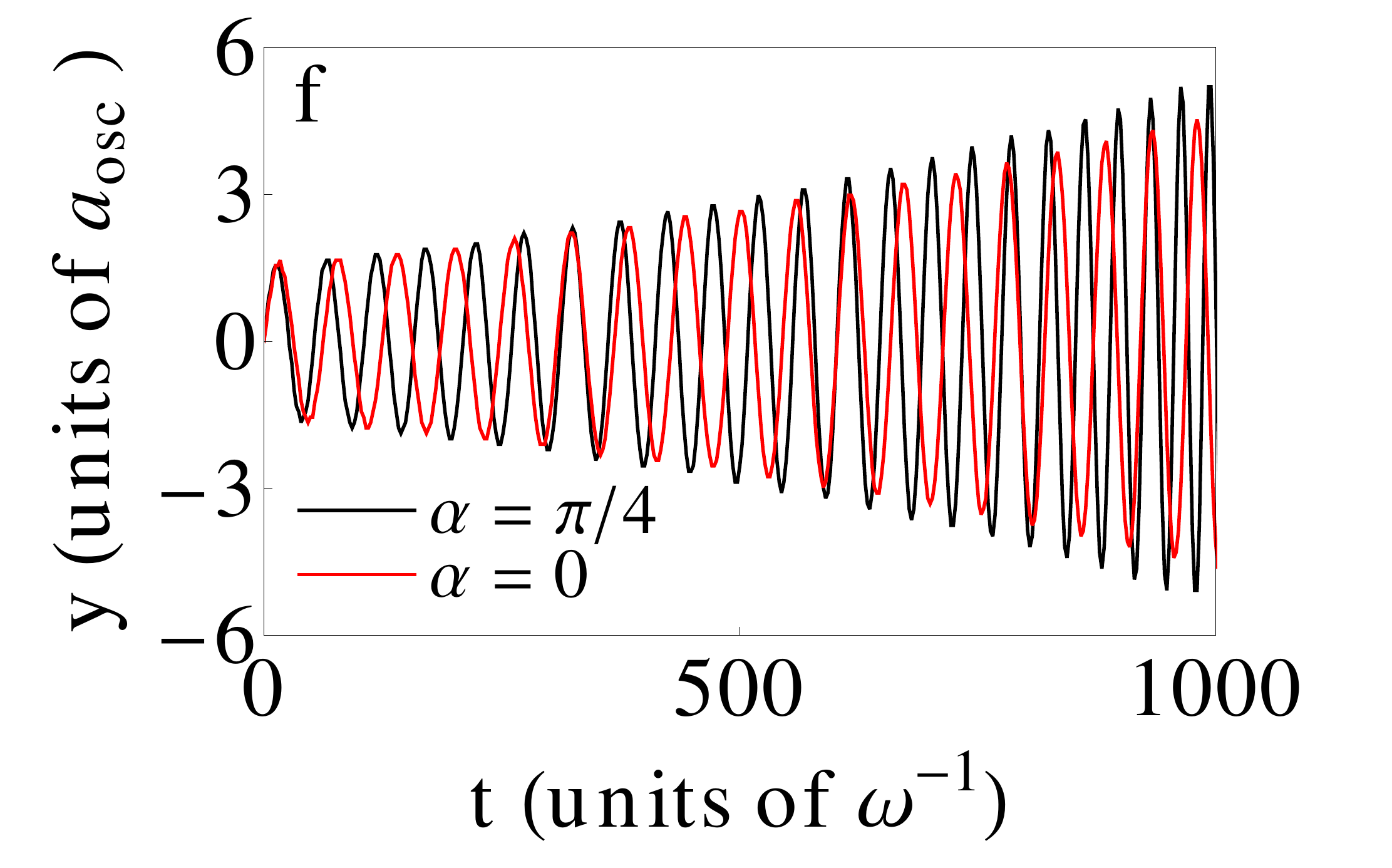}}
\end{tabular}
\caption{The top row shows the density profiles of the condensate with
a single vortex located at $(1.55~a_{\rm osc},0)$. The middle row shows 
the corresponding phase profiles. The angle $\alpha$, defining the 
direction of polarization, is equal to $0$ for (a) and (c) and 
$\pi/4$ for (b) and (d). The bottom row shows the dynamics
of the vortex in the two cases. The red and black curves correspond to
$\alpha = 0$ and $\alpha = \pi/4$ respectively.
}
\label{fig-1}
\end{figure}
\end{center}

\begin{center}
\begin{figure}[!h]
\begin{tabular}{c}
\resizebox{!}{!}
{\includegraphics[trim = 0cm 0cm 0cm 0cm,clip, angle=0,width=4cm]
                  {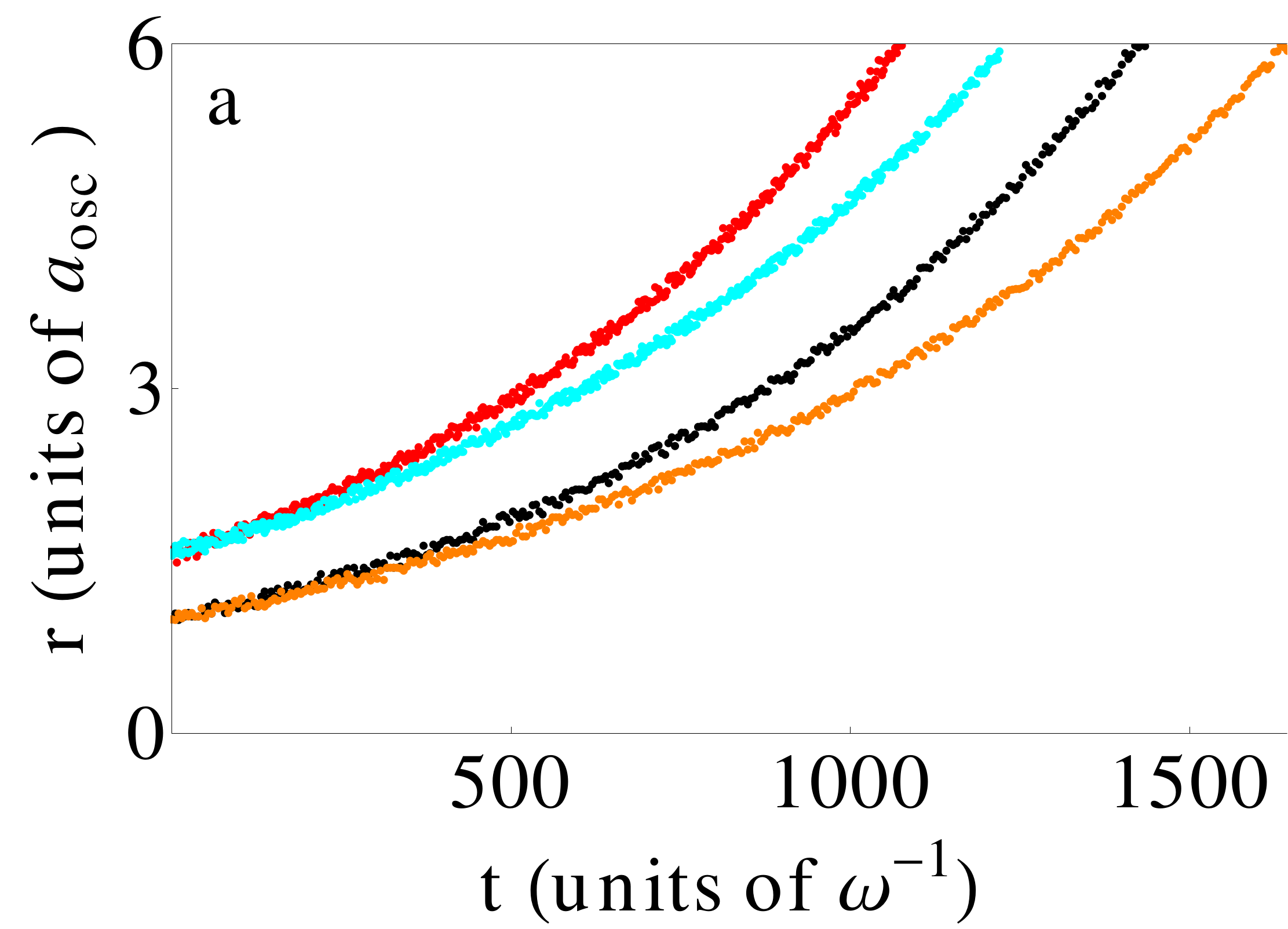}}
\resizebox{!}{!}
{\includegraphics[trim = 0cm 0cm 0cm 0cm,clip, angle=0,width=4.5cm]
                 {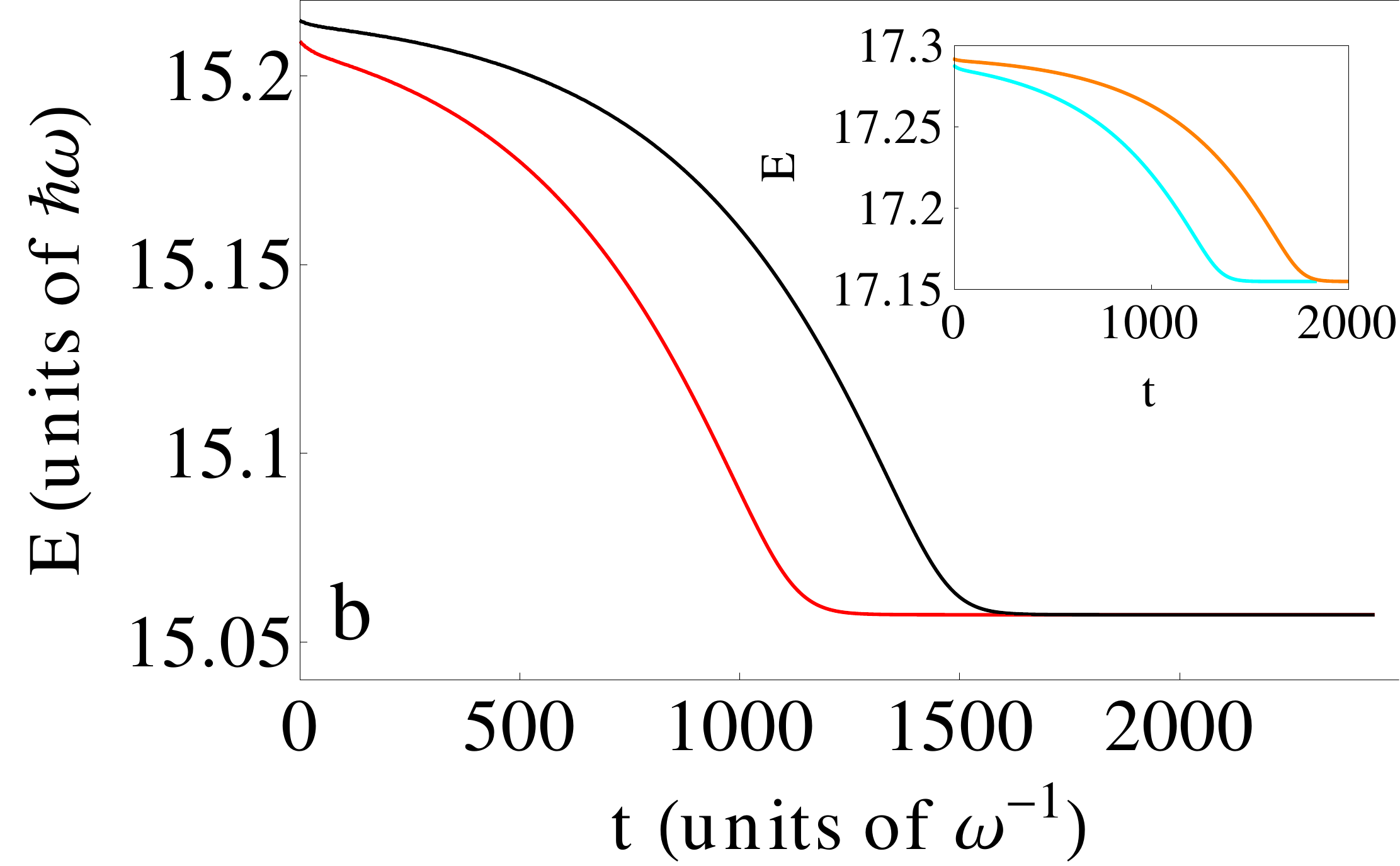}}
\end{tabular}
\caption{The left figure shows the dynamics of single vortex in a DBEC 
of $^{52}$Cr with $a = 50~a_0$, $a_{dd} = 16~a_0$, whereas the figure on
right shows the corresponding variation in the energy of the condensate. 
The angle $\alpha = 0$ for
the cyan and orange curves, whereas $\alpha=\pi/4$ for red and black curves. 
The upper and lower sets of curves correspond to two different initial locations 
of the vortices.
}
\label{fig-2}
\end{figure}
\end{center}

{\em Dynamics of a single vortex:} We consider $1.5\times10^5$ atoms $^{52}$Cr 
trapped in a harmonic trapping potential with 
$\omega_x = \omega_y  = 2\pi\times10$ Hz and $\omega_z = 2\pi\times100$ Hz. The 
back ground $s$-wave scattering length of $^{52}$Cr is $100~a_0$, where as 
dipolar length is $16~a_0$ \cite{Lahaye-2}. In order to accentuate the effect 
of the dipolar interaction, we consider $a = 50~a_0$ in the present work. In 
experiments, it can be achieved by tuning $s$-wave scattering length $a$ 
using magnetic Feshbach resonance \cite{Lahaye-1}. The unit of length and 
time used in the present work are thus $a_{\rm osc} = 4.41~\mu$m and 
$\omega^{-1} = 1.59\times10^{-2}$s respectively.
In order to generate the vortex in the system located at say 
$(x_0,y_0)$, we solve Eq.~(\ref{gpe_scaled}) in imaginary time $\tau = -it$ in 
the absence of any dissipation. After each iteration in imaginary time, we 
imprint the phase corresponding to the presence of the vortex, i.e., we 
perform the transformation 
\begin{equation}
\psi(\tau+\delta\tau) = \exp\left(i\tan^{-1}\frac{y-y_0}{x-x_0}\right)
                        |\psi(\tau)|,
\end{equation}
where $\delta\tau$ is the increment in the imaginary time. The solutions 
obtained by this method for two different directions of 
polarization, $\alpha = 0$ and $\alpha = \pi/4$, are shown in the upper 
two rows of Fig.~\ref{fig-1}. In both cases, the vortex is imprinted at 
$(1.55~a_{\rm osc},0)$.
It is evident from the Figs.~\ref{fig-1}(a) and (b) that for 
$\alpha = \pi/4$ the 
peak density has increased, whereas the extent of the density profile
has slightly decreased as compared to $\alpha = 0$ case. This is due
to the fact that for $\alpha = 0$ the dipolar interactions are purely
repulsive, whereas for $\alpha = \pi/4$ the component of the dipoles
along the $x$-axis interact attractively. Hence, the total dipolar
interaction, which is still repulsive, is nevertheless weaker for
$\alpha = \pi/4$. This leads to the decrease in the condensate size
and increase in the peak intensity. In order to study the dynamics of 
the vortex, we evolve these solutions in real time with non-zero 
dissipation. In the present work, we consider $\gamma = 0.0023$ which 
is reasonable choice of the dissipation \cite{Yan}. We find that the 
vortex slowly spirals out of the condensate (see the bottom panel of 
Fig.~\ref{fig-1}) as is the case for the condensate interacting via pure
contact interactions. This is not surprising as  
the dipolar interactions in a pancake-shaped DBEC with $\alpha = 0$ reduce 
to pure contact-like form \cite{Parker}. And, the DBEC is equivalent to the 
condensate interacting by pure contact interactions characterized by $s$-wave 
scattering length equal to $a+2a_{\rm dd}$, where $a$ and $a_{\rm dd}$ are 
the $s$-wave scattering and dipolar lengths of the DBEC. 
Now, the frequency of rotation of the vortex for 
$\alpha = \pi/4$ is greater than for $\alpha = 0$. This is due to the 
fact that in a homogeneous superfluid bounded by a circular boundary, 
the frequency of rotation of the point vortex is \cite{Alekseenko}
\begin{equation}
f = \frac{1}{4\pi^2(R^2 - r_0^2)},
\label{freq}
\end{equation}
where $R$ is the radius of the circular region and $r_0$ is the position
of the vortex. Approximating the radius of the circular region by
the root mean square radius of the condensate 
($R = \sqrt{\int r^2|\psi(\mathbf r)|^2dxdy}$), the ratio of rotation
frequencies of the vortices initially located at $(1.55,0)$ for 
$\alpha = \pi/4$ to $\alpha = 0$ is $1.18$ and is 
in very good agreement with the value $1.19$ obtained from the first
cycle in the bottom panel of Fig.~\ref{fig-1}. As the vortex moves
away from the center of the trap, the frequency increases due to the
decrease in the denominator of Eq.~(\ref{freq}), and this is consistent 
with the dynamics shown in bottom panel of Fig.~\ref{fig-1}, where the 
adjacent peaks start coming closer to each other with the progress of time.
For the same initial position of the vortex, the
rate of decay of the vortex is more for $\alpha = \pi/4$
as is shown in Fig.~\ref{fig-2}(a), where the radial
coordinate of the vortex is plotted as a function of time.
In order to calculate the the decay rate, we fit $r=r_0\exp(\gamma_d t)$
over the numerically calculated radial coordinate of the vortex, where
$\gamma_d$ is decay rate of the vortex initially located at $r_0$.
For the vortex initially located at $r_0 = 1.55~a_{\rm osc}$, the decay
rates obtained from this fitting are $1.263\times10^{-3}$ and $1.098 
\times10^{-3}$ for $\alpha = \pi/4$ and $\alpha = 0$ respectively. For the 
vortex initially located at $r_0 = 1.0~a_{\rm osc}$, the decay rates are 
slightly lower, $\gamma_d = 1.261\times10^{-3}$ for $\alpha = \pi/4$ 
and $\gamma_d = 1.094\times10^{-3}$ for $\alpha = 0$. Hence the decay 
rate is suppressed, although marginally, as 
the vortex moves nearer to the center of the trap.
The suppression of the decay rate as the vortex moves closer to the
trap center is also observed in condensates interacting via contact
interactions \cite{Rooney, Allen}. In the absence of dissipation
at $T = 0$ K, an off-center vortex traverses along an isoenergetic 
circular trajectory around the trap center due to the excitation 
of the anomalous or fundamental Kelvin mode \cite{Dodd,Simula, Middelkamp_2}. 
The non-zero dissipation at finite temperatures slowly reduces the energy of 
the condensate, which is shown in Fig.~\ref{fig-2}(b), by making
the vortex execute a spiral trajectory towards the edge of the condensate 
\cite{Rokhsar,Svidzinsky}. This leads to a vortex-free state which now is 
energetically stable.

\begin{center}
\begin{figure}[!h]
\begin{tabular}{c}
\resizebox{!}{!}
{\includegraphics[trim = 1.5cm 0cm 1.5cm 0cm,clip, angle=0,width=4cm]
                 {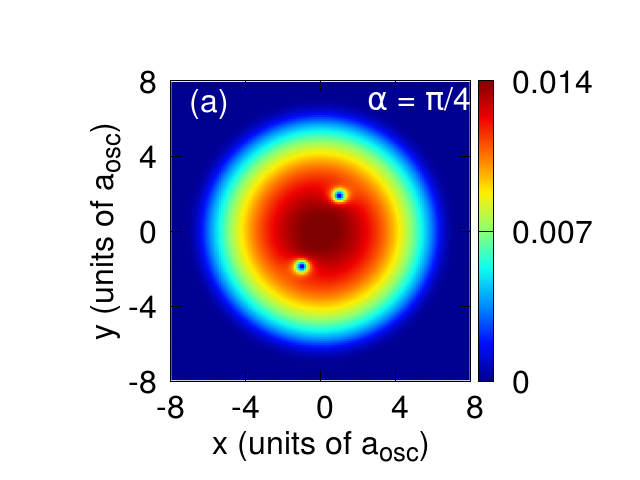}}
\resizebox{!}{!}
{\includegraphics[trim = 1.5cm 0cm 1.5cm 0cm,clip, angle=0,width=4cm]
                 {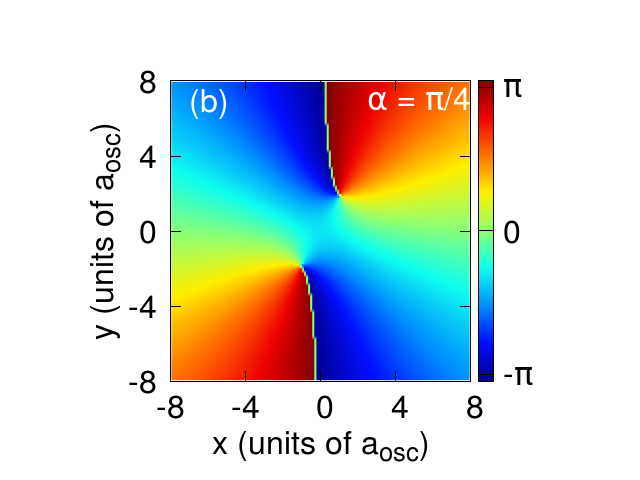}}
\end{tabular}
\caption{(a) shows the density and (b) shows the phase of the DBEC 
 of $^{52}$Cr atoms with $a = 50~a_0$ and $a_{\rm dd} = 16~a_0$. The angle
 $\alpha = \pi/4$.}
\label{fig-3}
\end{figure}
\end{center}

\begin{center}
\begin{figure}[!h]
\begin{tabular}{c}
\resizebox{!}{!}

{\includegraphics[trim = 0cm 0mm 0cm 0mm,clip, angle=0,width=4cm]
                         {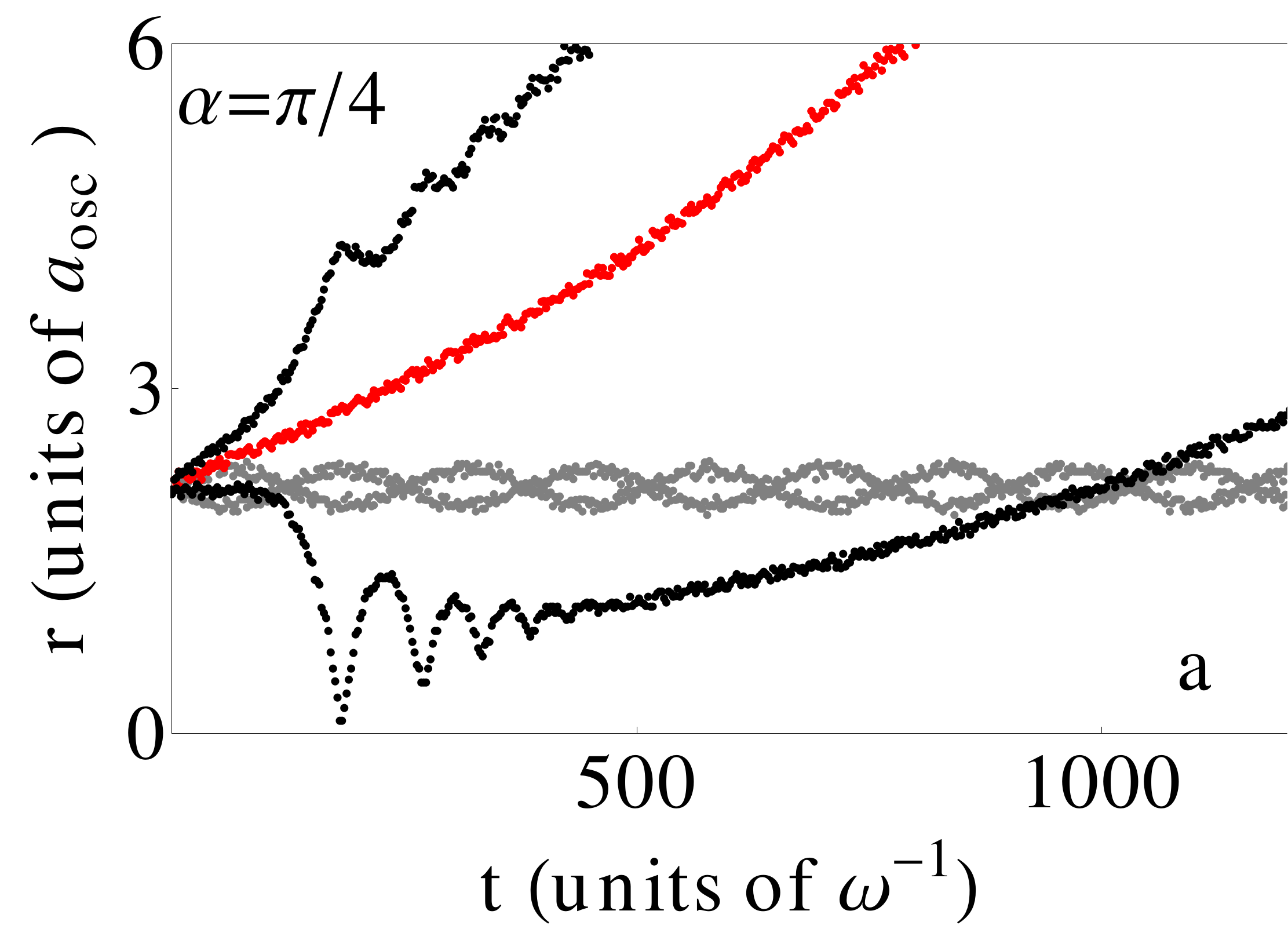}}
{\includegraphics[trim = 0cm 0mm 0cm 0mm,clip, angle=0,width=4.2cm]
                         {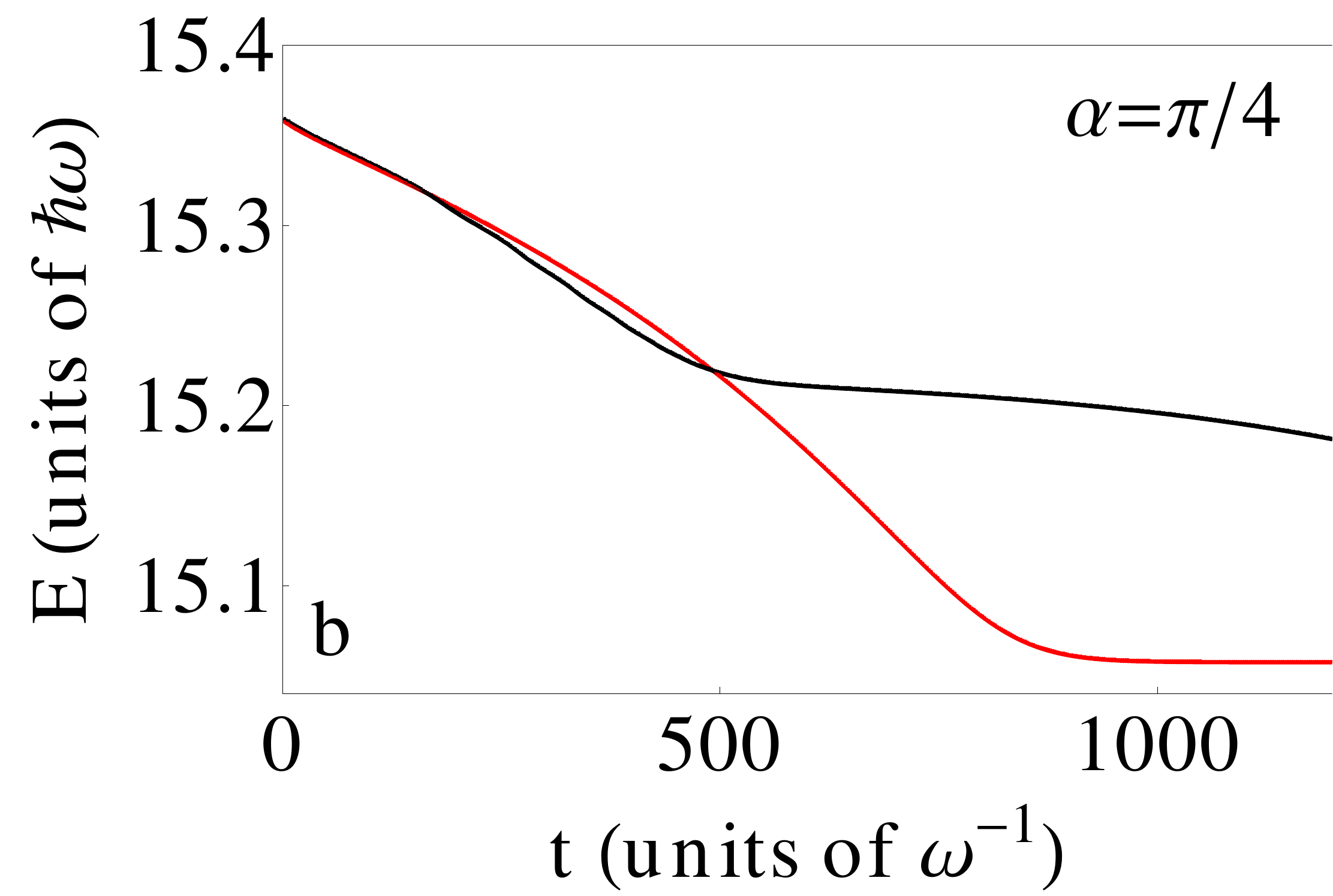}}
\end{tabular}
\caption{The left figure shows the radial coordinates of the two corotating 
vortices as a function of time in a DBEC of $^{52}$Cr with
$a = 50~a_0$, $a_{dd} = 16~a_0$. The right figure shows the corresponding
variation in the energy of the condensate with time. Red dots and line 
correspond to the vortices which are symmetrically located about origin at 
equal radial distances, whereas for the black dots and line there is a slight 
asymmetry in the locations of the two vortices. The value of dissipation $\gamma = 0.0023$,
except for the gray dots in the left figure, which show the radial coordinates 
of the asymmetrically located vortices as a function of time in the absence of 
dissipation.
}
\label{fig-4}
\end{figure}
\end{center}

\begin{center}
\begin{figure}[!h]
\begin{tabular}{c}
\resizebox{!}{!}
{\includegraphics[trim = 0cm 0cm 0cm 0cm,clip, angle=0,width=4cm]
                  {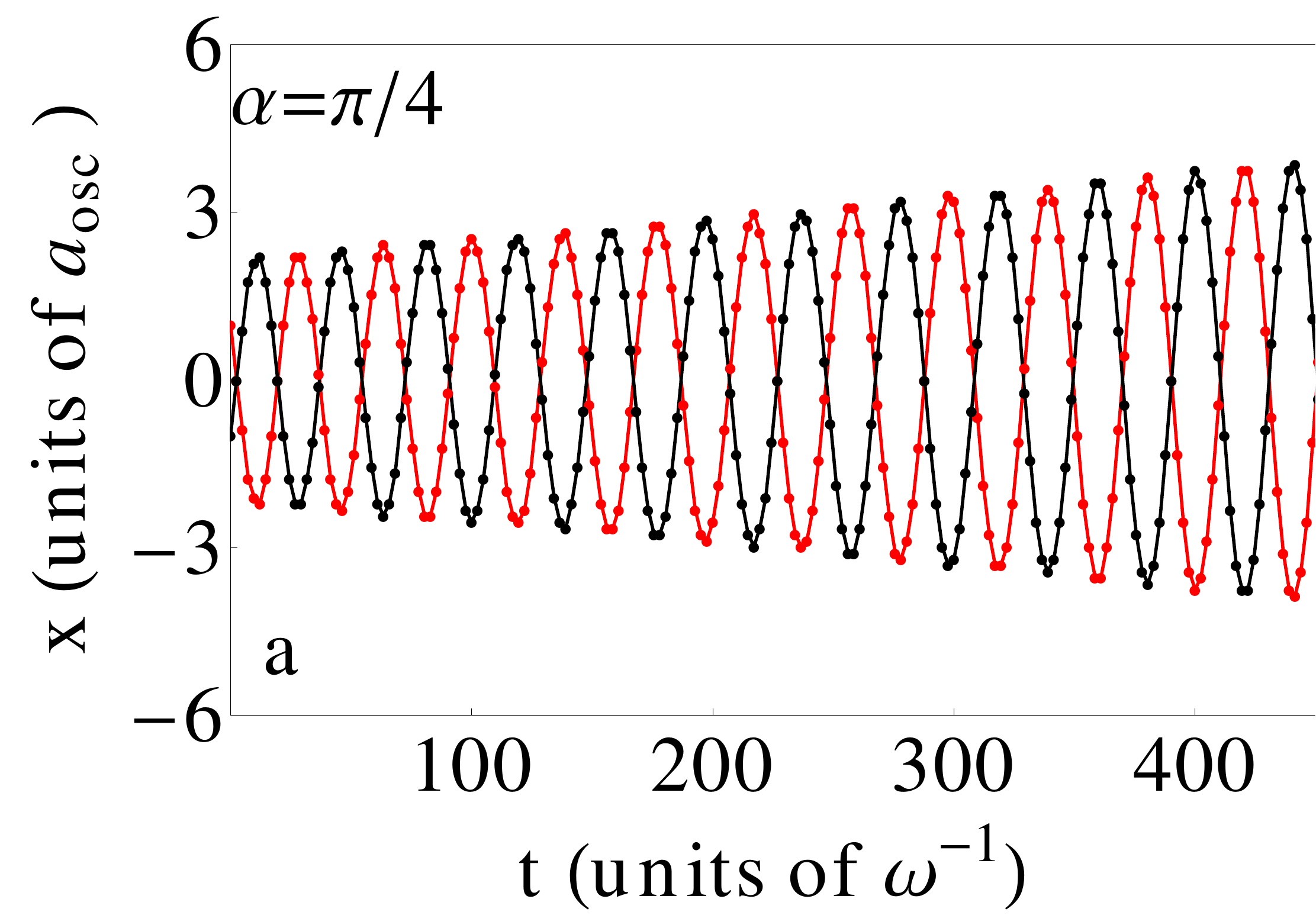}}
\resizebox{!}{!}
{\includegraphics[trim = 0cm 0cm 0cm 0cm,clip, angle=0,width=4cm]
                 {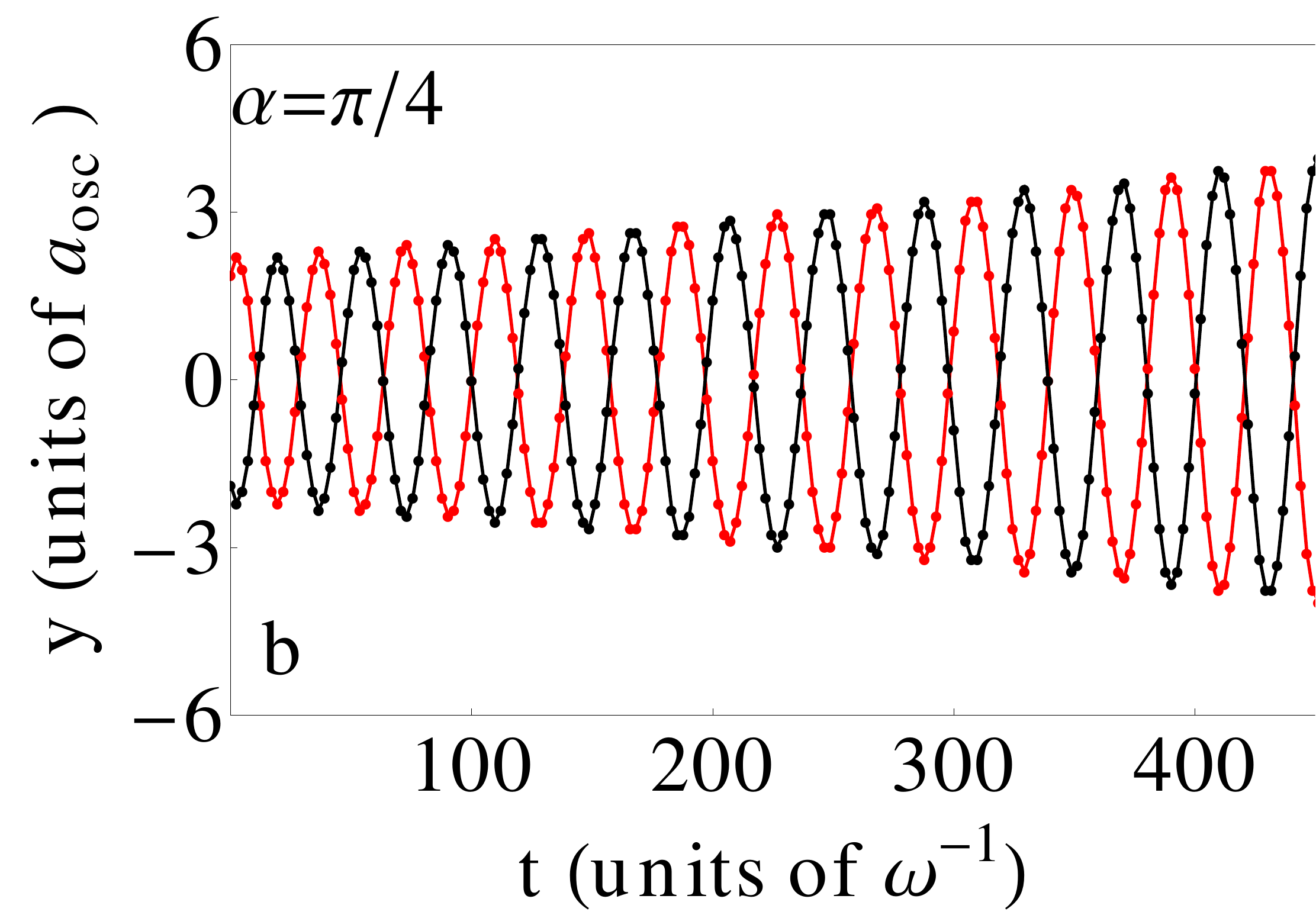}}\\
{\includegraphics[trim = 0cm 0cm 0cm 0cm,clip, angle=0,width=4cm]
                  {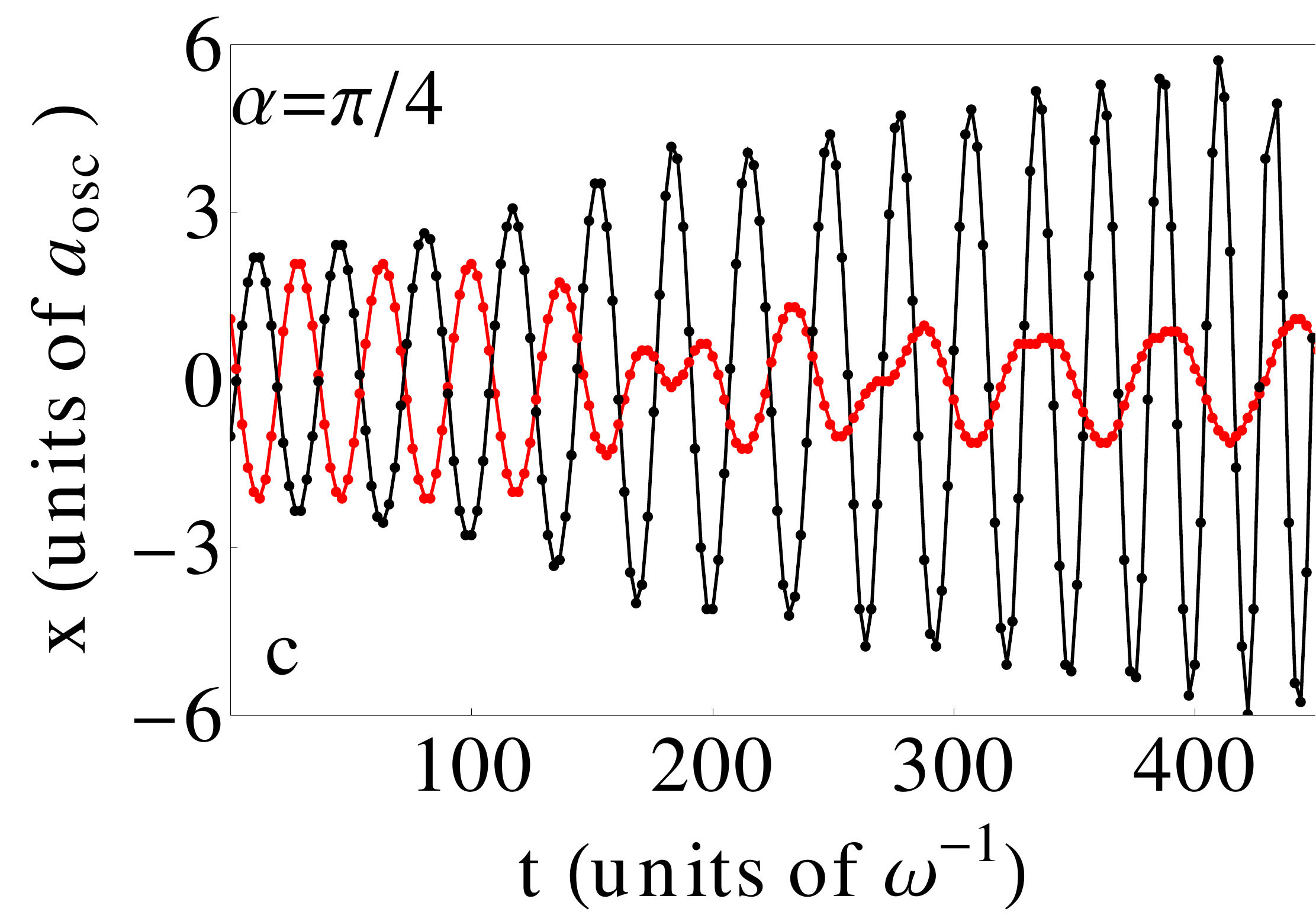}}
\resizebox{!}{!}
{\includegraphics[trim = 0cm 0cm 0cm 0cm,clip, angle=0,width=4cm]
                 {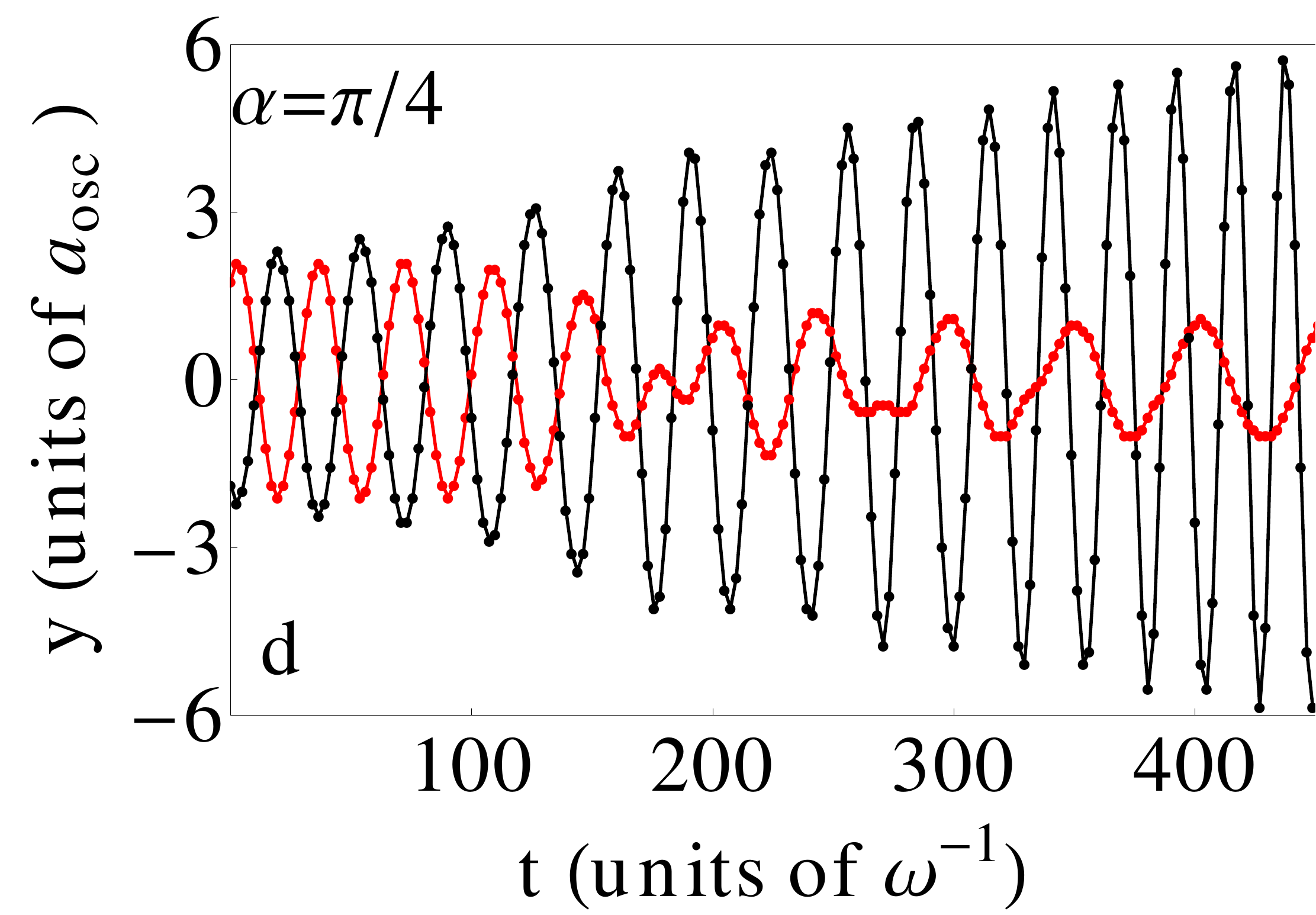}}\\
 \resizebox{!}{!}
{\includegraphics[trim = 0cm 0cm 0cm 0cm,clip, angle=0,width=4cm]
                  {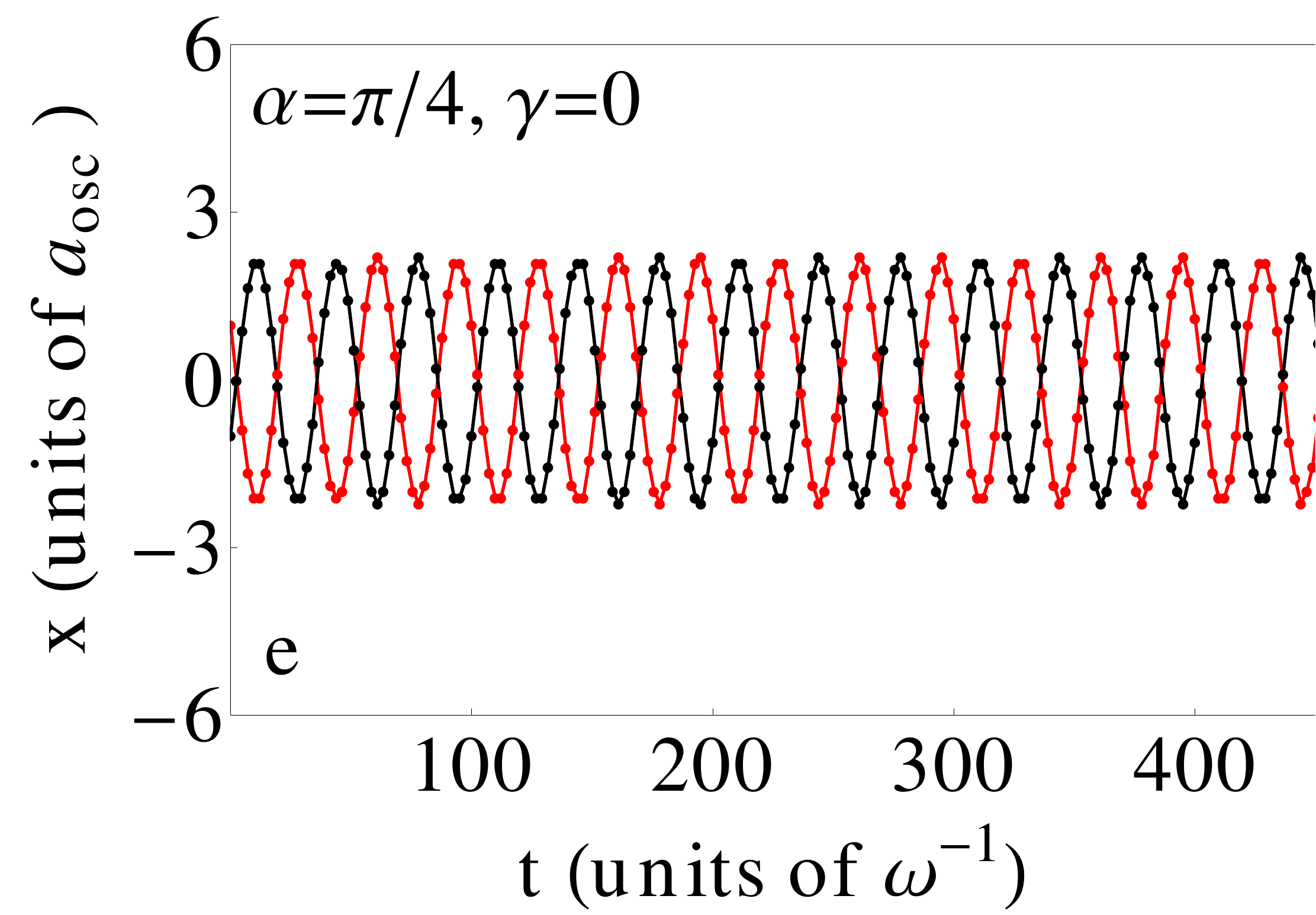}}
\resizebox{!}{!}
{\includegraphics[trim = 0cm 0cm 0cm 0cm,clip, angle=0,width=4cm]
                 {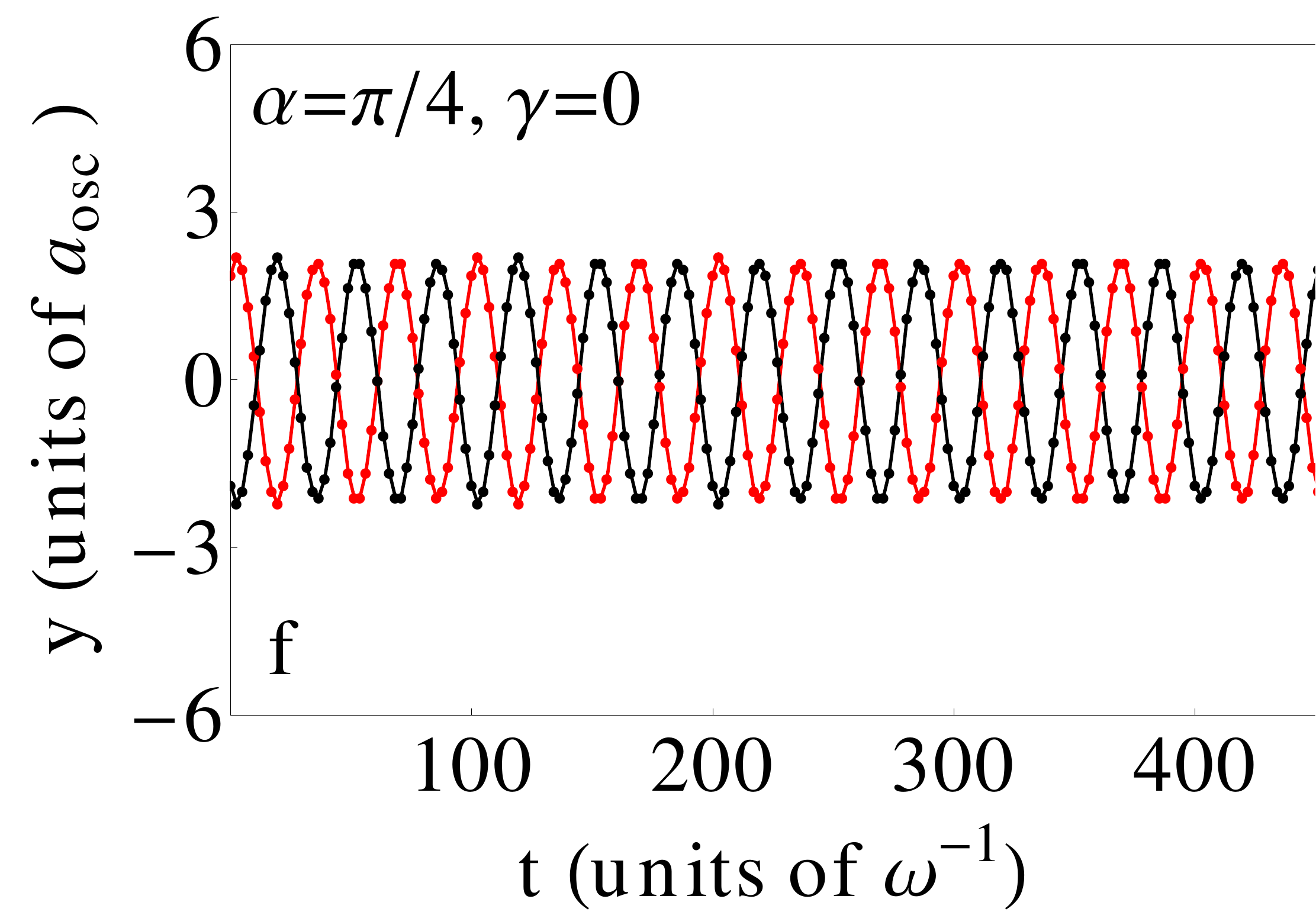}}\\
 \resizebox{!}{!}
{\includegraphics[trim = 0cm 0cm 0cm 0cm,clip, angle=0,width=4cm]
                  {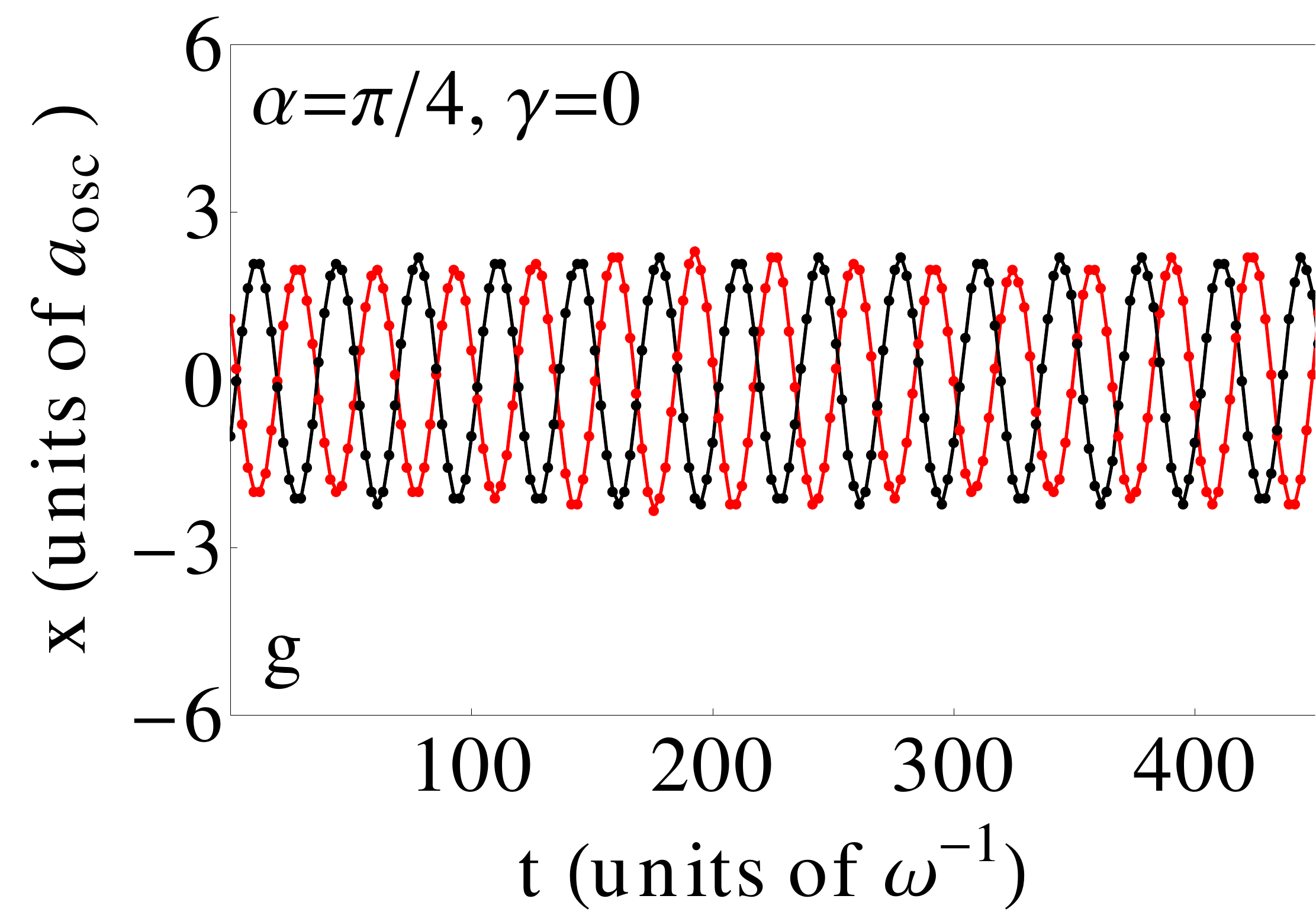}}
\resizebox{!}{!}
{\includegraphics[trim = 0cm 0cm 0cm 0cm,clip, angle=0,width=4cm]
                 {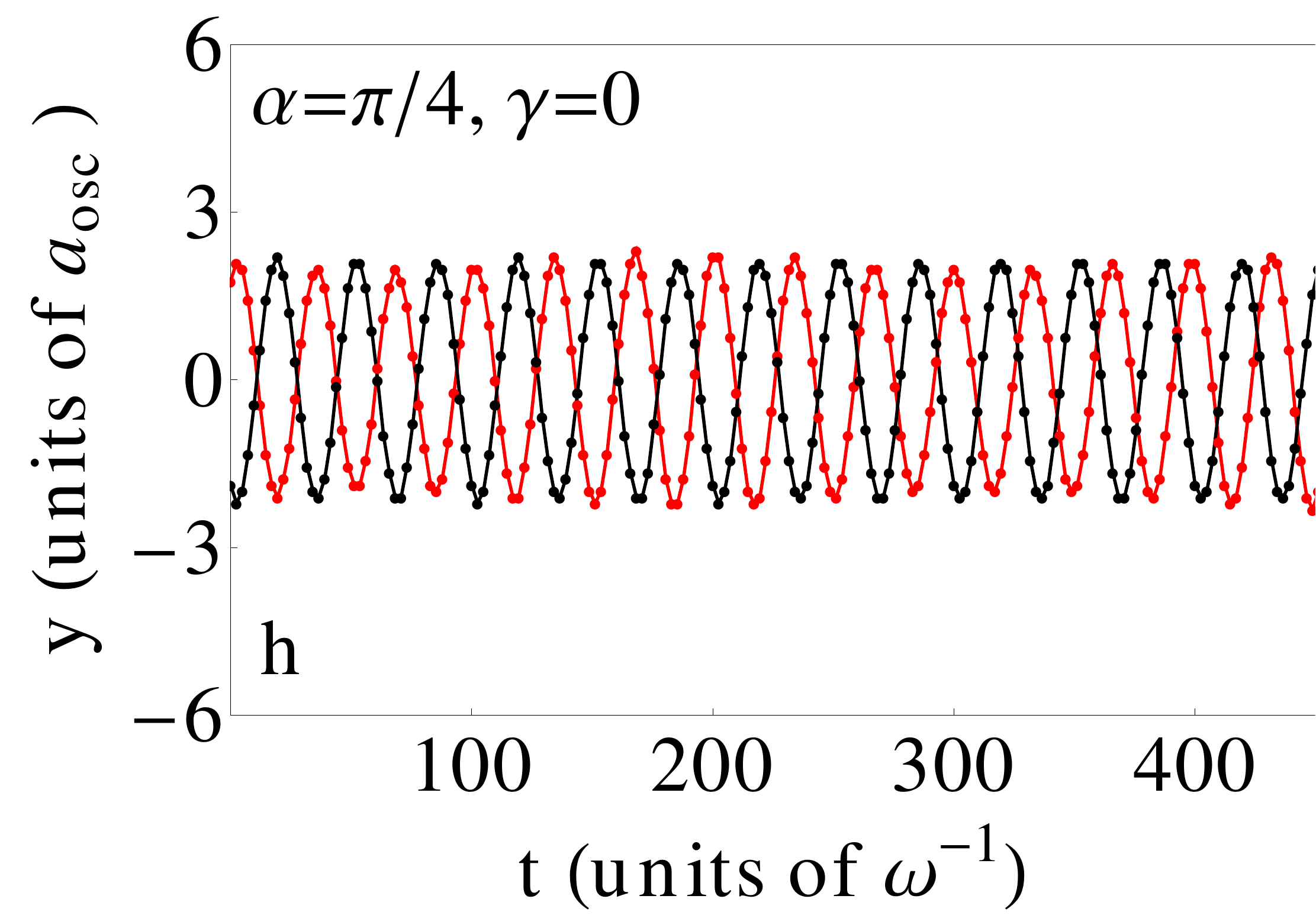}}                
\end{tabular}
\caption{ The first row shows the dynamics of the two corotating vortices
initially located at $(-1.0~a_{\rm osc},-1.88~a_{\rm osc})$
and $(1.0~a_{\rm osc},1.88~a_{\rm osc})$ at finite $T$, whereas third row 
shows the dynamics of the pair at $T = 0$ K. The second row shows the dynamics 
of the two corotating vortices initially located at 
$(-1.0~a_{\rm osc},-1.88~a_{\rm osc})$ and 
$(1.1~a_{\rm osc},1.77~a_{\rm osc})$ at finite $T$, whereas the fourth row
shows the dynamics of the pair at $T = 0$ K.
}
\label{fig-5}
\end{figure}
\end{center}

\begin{center}
\begin{figure}[!h]
\includegraphics[trim = 0cm 0mm 0cm 0mm,clip, angle=0,width=7.5cm]
                 {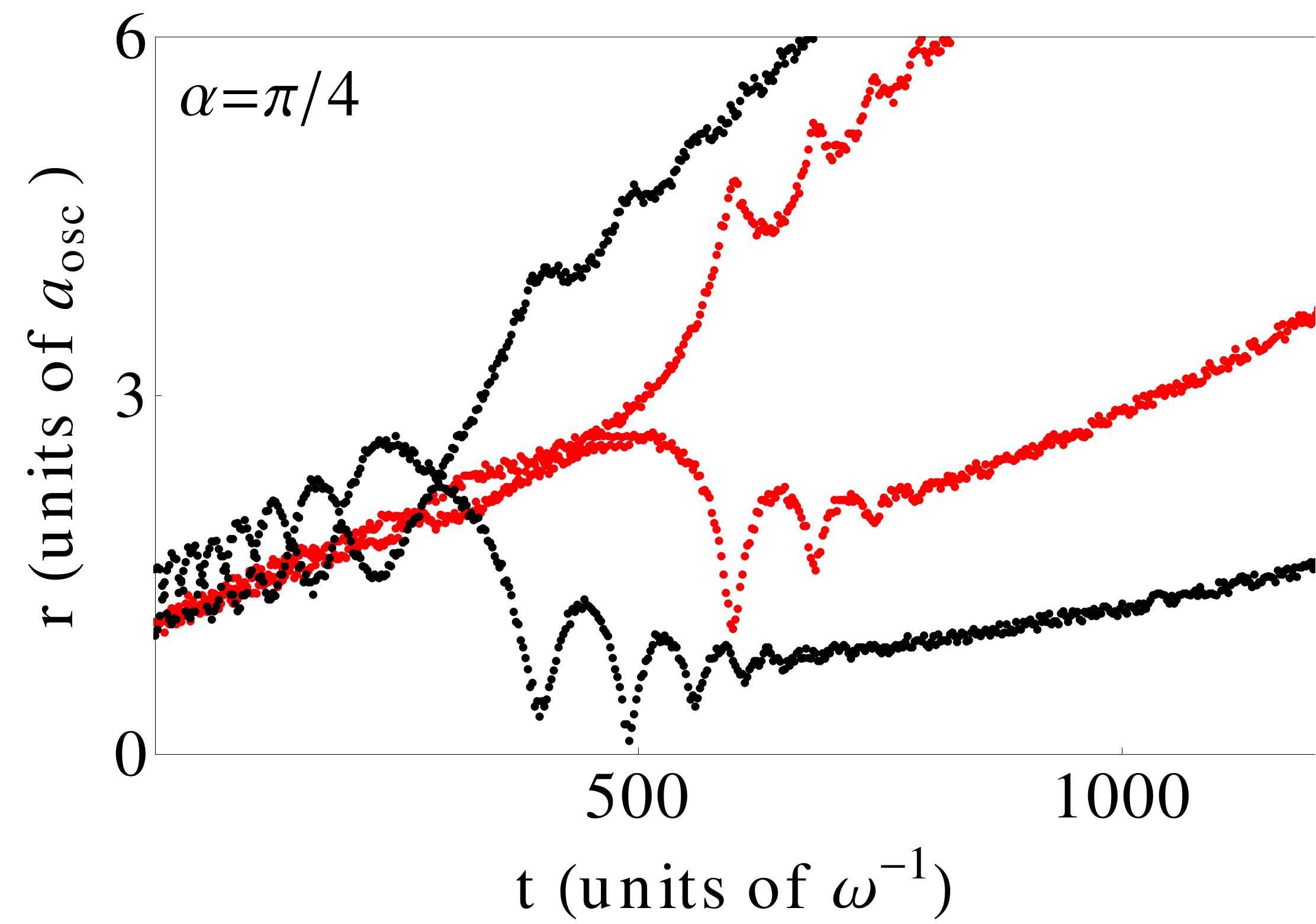}
\caption{Red dots are the radial coordinates of the corotating vortices
initially located at $(-1.1,0.0)$, $(1.0,0.0)$. The black dots are the radial
coordinates of the corotating vortices initially located at 
$(-1.55,0.0)$, $(1.0,0.0)$. The period during which the two black 
(red) dotted curves cross each other is qualitatively similar to the zero 
temperature dynamics as shown by the gray dots in Fig.~\ref{fig-4}(a).}
\label{fig-6}
\end{figure}
\end{center}

\begin{center}
\begin{figure}[!h]
\begin{tabular}{c}
\resizebox{!}{!}
{\includegraphics[trim = 0cm 0cm 0cm 0cm,clip, angle=0,width=8.5cm]
                  {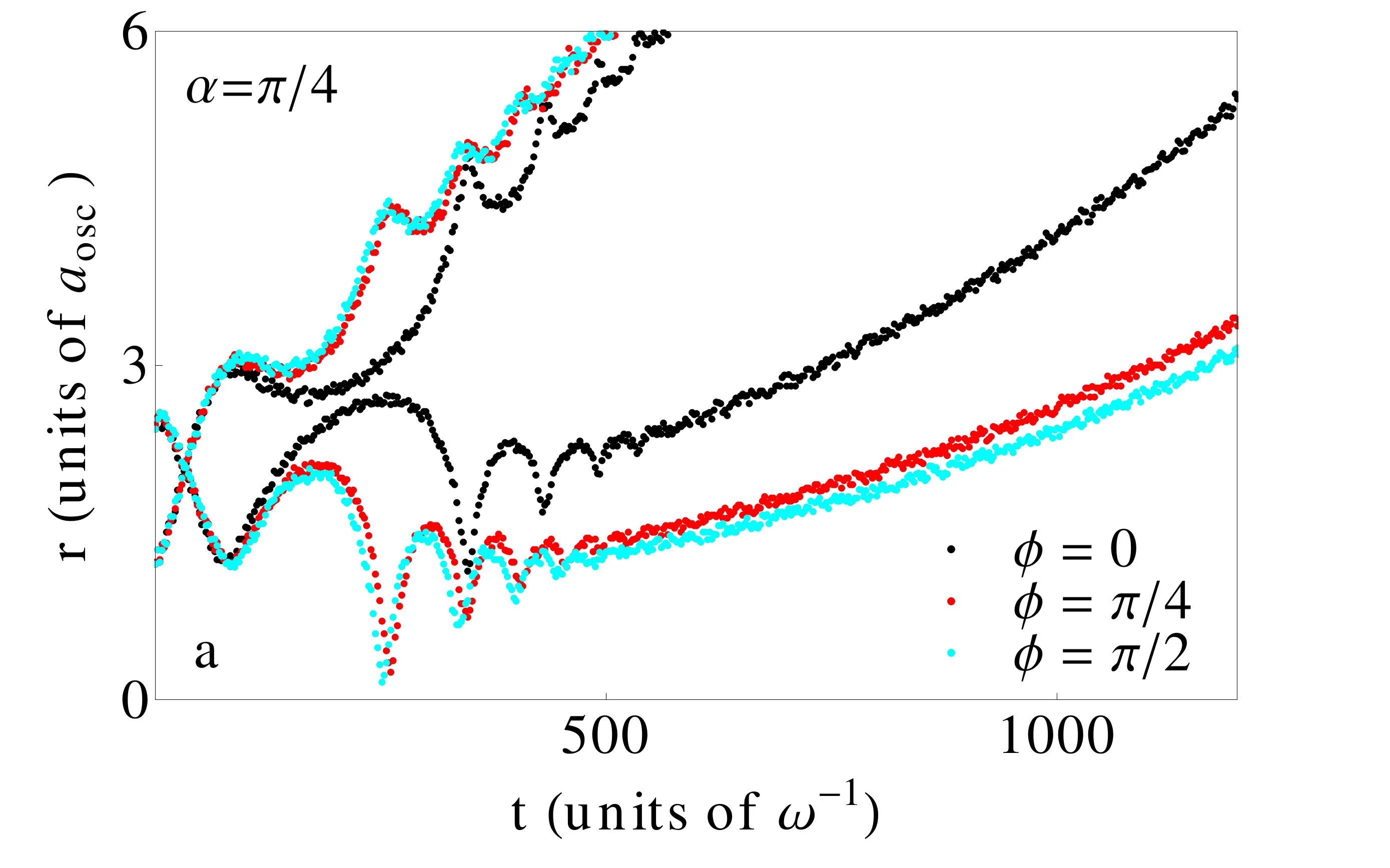}}\\
\resizebox{!}{!}
{\includegraphics[trim = 0cm 0cm 0cm 0cm,clip, angle=0,width=8.5cm]
                 {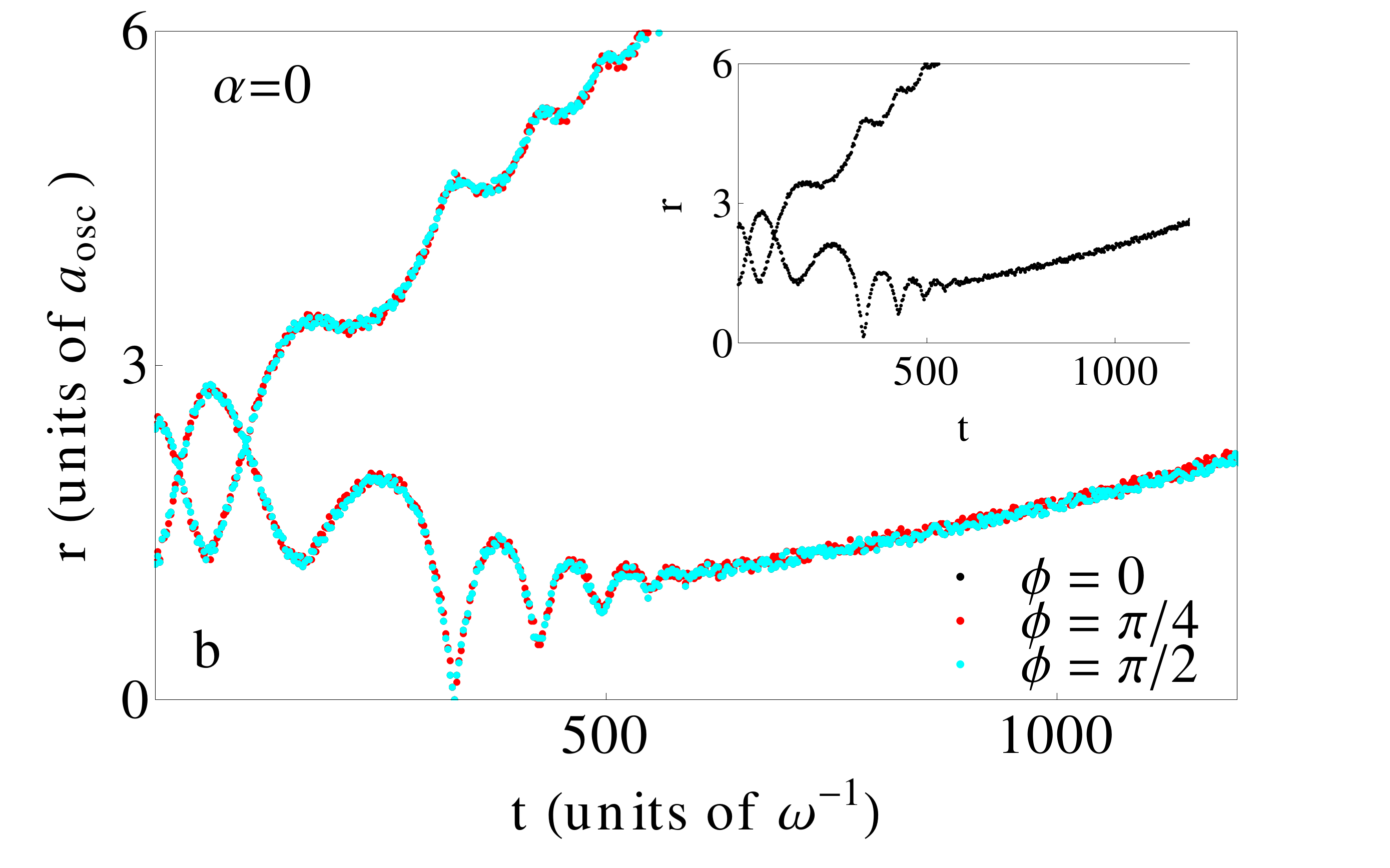}}
\end{tabular}
\caption{The dynamics of a corotating vortex pair in a DBEC of $1.5\times10^5$ $^{52}$Cr 
 atoms with $a = 50~a_0$, $a_{dd} = 16~a_0$ for three different initial
 orientations. The angle $\alpha=\pi/4$ for (a) and $\alpha = 0$ for (b). 
 The angle $\phi$ is equal to $0$, $\pi/4$, and $\pi/2$ for black, red, and 
 cyan dots respectively. The inset in (b) shows the dynamics of the 
 two corotating vortices which are located diametrically opposite to each 
 other at radial coordinates $r_1 = \sqrt{1.1^2+2.21^2}$, $r_2 = \sqrt{0.5^2+1.1^2}$ 
 and arbitrary $\phi$ in a condensate of $^{52}$Cr atoms with $a = 82~a_0$ and 
 $a_{dd} = 0$.
 }
\label{fig-7}
\end{figure}
\end{center}

{\em Dynamics of a corotating vortex pair:} In order to study the dynamics 
of two corotating vortices, we first generate the minimum energy solution 
by solving Eq.~(\ref{gpe_scaled}) in imaginary time $\tau$ in the absence 
of any dissipation. As in the case of a single vortex, we transform $\psi$ 
after each iteration in time as
\begin{eqnarray}
\psi(\tau+\delta\tau) &=& \exp\left\{i\left[\tan^{-1}\left(\frac{y-y_1}{x-x_1}\right)\right.\right.\nonumber\\
&&\left.\left.+\tan^{-1}\left(\frac{y-y_2}{x-x_2}\right)\right]\right\}|\psi(\tau)|,
\end{eqnarray}
here $(x_1,y_1)$ and $(x_2,y_2)$ are the locations of the two vortices.
The solution obtained by this method for the two vortices initially 
located at $(-1.0~a_{\rm osc},-1.88~a_{\rm osc})$ and 
$(1.0~a_{\rm osc},1.88~a_{\rm osc})$ is shown in Fig.~\ref{fig-3}.
This minimum energy solution is now evolved in real time in the presence
of dissipation. In the case of a corotating vortex pair, the dynamics 
depends upon the initial asymmetry in the location of the vortices. 
For the two
vortices symmetrically located about origin, i.e. $x_2= -x_1$ and
$y_2=-y_1$ as is the case for the solution shown in Fig.~\ref{fig-3}, 
both the vortices decay at the same rate and slowly spiral
out of the condensate like a single vortex as is shown by red dots in 
Fig.~\ref{fig-4}(a); the full dynamics is shown in 
Fig.~\ref{fig-5}(a) and (b). The corresponding variation in the energy of the
system is shown by the red curve in Fig.~\ref{fig-4}(b). In this case,
we can define the decay rate as in the case of a single vortex by
fitting $r = r_0\exp(\gamma_d t)$ over the numerical values of radial
coordinates. The decay rate thus obtained is $1.35\times10^{-3}$. 
In case the vortices are not symmetrically located about origin, the dynamics
becomes qualitatively different. Here the two vortices don't decay at the
same rate. Instead of this, the decay rate of one of the vortex is suppressed
which starts moving closer to the center of the trap, whereas the decay rate
of the second vortex is increased as is shown by black dots in 
Fig.~\ref{fig-4}(a). The initial coordinates of the vortices in this case are  
$(-1.0~a_{\rm osc},-1.88~a_{\rm osc})$ and 
$(1.1~a_{\rm osc},1.77~a_{\rm osc})$. If one examines the variation in
energy in this case, shown by the black curve in Fig.~\ref{fig-4}(b), there
are two distinct regions. For the region $t\lessapprox490~\omega^{-1}$, energy 
of the asymmetric vortex pair decays faster than the symmetric vortex pair.
This is due to the increase in the separation between the two 
vortices leading to the decrease in the inter-vortex interaction energy. 
The faster decaying vortex ultimately exits out of the condensate leaving a 
single vortex, which now decays as a single vortex after 
$t\gtrapprox490~\omega^{-1}$. The full dynamics for the symmetrically and 
asymmetrically located pair of vortices is shown in Figs.~\ref{fig-5}(a-d).
The finite temperature dynamics is in significant contrast to the dynamics at 
$T = 0$ Kelvin $(\gamma = 0)$, which is shown in Figs.~\ref{fig-5}(e-h).
The radial coordinates of the asymmetrically located vortices at $T = 0$
as a function of time are shown by grey dots in Fig.~\ref{fig-4}(a). It is
evident that there are oscillations in the radial coordinates of 
the asymmetrically located vortices at $T = 0$ K. 
From the Figs.~\ref{fig-4}(a) [the dynamics depicted by black dots], \ref{fig-5}(c),
and  \ref{fig-5}(d), it is evident that from the very beginning, one of 
the vortex starts moving towards the center, while the other starts moving 
towards the edge of the condensate. This, however, is not the case always. 
We find that for a different set of initial vortex locations, the component 
vortices may alternatively end up nearer to the trap center, especially during 
the initial stages of the dynamics. This is evident from the black dots in 
Fig.~\ref{fig-6}, where the dynamics of two vortices initially located at 
$(-1.55~a_{\rm osc},0.0)$, $(1.0~a_{\rm osc},0.0)$ is shown. Even when the 
initial asymmetry is 
decreased, there are radial oscillations (see red dots in Fig.~\ref{fig-6}). 
These radial oscillations are reminiscent of the dynamics of asymmetrically 
located vortices at zero temperature, see the dynamics depicted by gray dots
in Fig.~\ref{fig-4}(a), \cite{Gautam}. 
Due to this one of the vortices ends up decaying faster than the other. 
At finite temperature, the random fluctuations will invariably lead to the 
asymmetry in locations of the component vortices \cite{Gautam}.
{\em Effect of anisotropic vortex-vortex interactions on the dynamics:} 
It was pointed out in Ref.~\cite{Mulkerin} that if the magnetic dipoles have 
non-zero component on the plane of the condensate, the interactions between 
the vortices become anisotropic. In order to study the effect of anisotropic 
interactions on the dynamics of corotating vortices, we consider the dynamics 
of the vortex pair for three different initial locations. These are
(a) $(-\sqrt{1.1^2+2.21^2},0)$, $(\sqrt{0.5^2+1.1^2},0)$; (b) $(-1.1,-2.21)$,  
$(0.55, 1.1)$; and (c) $(0, -\sqrt{1.1^2+2.21^2})$, $(0, \sqrt{0.5^2+1.1^2})$ 
in scaled units. In these three cases, the components vortices have the same 
radial coordinates, and the relative position vector 
$\mathbf{r}_2-\mathbf{r}_1$ makes an angle $0$, $\pi/4$ or $\pi/2$ with the 
$x$-axis. Let us denote this angle by $\phi$. If $\alpha = \pi/4$, the dynamics
for these three cases is different as is shown in Fig.~\ref{fig-7}(a). This is 
due to different interaction energy between the two vortices for these three 
cases. The interaction energy between the two vortices is defined as 
\cite{Mulkerin}
\begin{equation}
E_{\rm int}(\mathbf{r}_2-\mathbf{r}_1) = E_2(\mathbf{r}_1,\mathbf{r}_2) - 
                               E_1(\mathbf{r}_1) - E_1(\mathbf{r}_2), 
\end{equation}
where $E_j$ with $j=1,2$ is the energy associated with $j$ vortices in the
system and is equal to total energy of the system minus the energy of
the vortex free state. The interaction energy thus calculated are 
$0.0232~\hbar\omega$ for (a),  
$0.0229~\hbar\omega$ for (b), and $0.0228~\hbar\omega$ for (c). We checked
the precision of these results upto four decimal places using two different 
spatial and imaginary time step sizes to solve the DGPE in imaginary 
time - (a) $\delta x = \delta y = 0.11~a_{\rm osc}$ and 
$\delta \tau = 0.0006$ and (b) $\delta x = \delta y = 0.055~a_{\rm osc}$ 
and $\delta \tau = 0.0002$. Hence the interaction energy is maximum
for the vortices located along $x$-axis and minimum when they are located along
$y$-axis in agreement with Ref.~\cite{Mulkerin}. This anisotropy in the vortex 
interaction leads to different dynamics in the aforementioned three cases.
In case $\alpha = 0$, the vortex-vortex interaction is isotropic and is 
manifested by same dynamics for the aforementioned three cases as is shown 
in Fig.~\ref{fig-7}(b). The initial vortex-vortex interaction energy in this
case is $0.0241~\hbar\omega$ irrespective of the value of $\phi$. 
As mentioned earlier, a pancake-shaped DBEC with 
$\alpha = 0$ is equivalent to a condensate interacting via pure
contact interactions. In order to compare
the dynamics of the corotating vortex pair in such a condensate with that of
the DBEC, we consider a condensate consisting of $1.5\times10^5$ atoms of $^{52}$Cr 
with $a = 82~a_0$ and $a_{\rm dd} = 0$.
The dynamics of two corotating vortices, 
initially located diametrically opposite to each other at radial coordinates 
$r_1 = \sqrt{1.1^2+2.21^2}$, $r_2 = \sqrt{0.5^2+1.1^2}$ and any arbitrary 
$\phi$ is shown in the inset of Fig.~\ref{fig-7}(b). It is evident that the
dynamics in two cases is same. This equivalence in dynamics of 
vortices in the pancake-shaped DBEC with $\alpha = 0$ and the condensate 
interacting with pure contact interactions does not depend upon $\gamma$, and 
hence is true even at $T = 0$ K $(\gamma = 0)$. 
\section{Summary of results}
\label{Sec-IV}
We have studied the dynamics of a single vortex and two corotating vortices in
the quasi two-dimensional dipolar condensates in the presence of dissipation.
A single vortex in the presence of dissipation spirals out of the condensate
due to the energetic instability. We observe that making the dipolar 
interactions partially attractive by changing the polarization direction, 
leads to the decrease in size of the condensate and faster decay rate for a 
single vortex. In the case of two symmetrically located corotating vortices, 
the finite temperature dynamics is qualitatively similar to the single vortex,
and the decay rate of the each vortex is equal. On the other hand, the 
introduction of the slight asymmetry in the initial locations leads to the
different dynamics. In this case, the decay rate of the one of the vortex
is suppressed at the cost of the other. For the dipoles not oriented normal
to the condensate plane, the vortex-vortex interactions are no longer 
isotropic. We observe that even a small difference in vortex-vortex interaction
energy can lead to the perceptible difference in the dynamics at finite
temperature.  

%%%%%%%%%%%%%%%%%%%%%%%%%%%%%%%%%%%%%%%%%%%%%%%%%%%%%%%%%%%%%%%%%%%%%%%%%%%%%%%
%%%%%%%%%%%%             Acknowledgements                          %%%%%%%%%%%%
%%%%%%%%%%%%%%%%%%%%%%%%%%%%%%%%%%%%%%%%%%%%%%%%%%%%%%%%%%%%%%%%%%%%%%%%%%%%%%%

\begin{acknowledgements}
We thank Prof. S.~K.~Adhikari for useful comments and suggestions.
This work is financed by Funda\c{c}\~ao de Amparo \`a Pesquisa do Estado de S\~ao Paulo 
(FAPESP) under Contract No. 2013/07213-0.
\end{acknowledgements}

%%%%%%%%%%%%%%%%%%%%%%%%%%%%%%%%%%%%%%%%%%%%%%%%%%%%%%%%%%%%%%%%%%%%%%%%%%%%%%%
%%%%                        Bibliography                                  %%%%%
%%%%%%%%%%%%%%%%%%%%%%%%%%%%%%%%%%%%%%%%%%%%%%%%%%%%%%%%%%%%%%%%%%%%%%%%%%%%%%%

\end{document}